\documentclass{jaa}
\usepackage{natbib}
\usepackage{hyperref}
\hypersetup{colorlinks=true, linkcolor=blue, filecolor=blue, citecolor = blue, urlcolor=cyan,} 
\usepackage{graphicx}
\usepackage{adjustbox} 
\usepackage{subcaption} 
\newcommand{\ie}{$i.e.,\;$}
\newcommand{\eg}{$e.g.,\;$}
\newcommand{\viz}{$viz.,\;$}

\newcommand{\wrt}{$w.r.t.,\;$}
\begin{document}\sloppy
%
\title{Detection of radio-AGN in dust-obscured galaxies using deep uGMRT radio continuum observations}
%
%
%
%
\author{Abhijit Kayal\textsuperscript{1,2*}, Veeresh Singh\textsuperscript{1}, C.H. Ishwara Chandra\textsuperscript{3}, Yogesh Wadadekar\textsuperscript{3} and Sushant Dutta\textsuperscript{1,2} }
\affilOne{\textsuperscript{1}Physical Research Laboratory, Ahmedabad, 380009, Gujarat, India.\\}
\affilTwo{\textsuperscript{2}Indian Institue of Technology Gandhinagar, Palaj, Gandhinagar, 382355, Gujarat, India.\\}
\affilThree{\textsuperscript{3}National Centre for Radio Astrophysics, TIFR, Post Bag 3, Ganeshkhind,Pune 411007, India.}
%
%
\twocolumn[{

\maketitle

\corres{abhijitk@prl.res.in}

\msinfo{24 March 2022}{6 June 2022}

\begin{abstract}
Radio observations being insensitive to the dust-obscuration, have been exploited to unveil the population of Active Galactic 
Nuclei (AGN) residing in galaxies with large dust content. 
In this paper, we investigate the radio characteristics of 321 dust-obscured galaxies 
(DOGs; S$_{\rm 24~{\mu}m}$/S$_{\rm r~band}$ $\geq$1000) by using mainly
deep band-3 (250$-$550 MHz) observations from the upgraded Giant Metrewave Radio Telescope (uGMRT) and 
1.5 GHz Jansky Very Large Array (JVLA) observations. We find that, for our sample of DOGs, deep (median noise-rms $=$ 30 $\mu$Jy beam$^{-1}$) 400 MHz band-3 uGMRT observations yield the highest detection rate (28 per cent) among those obtained with the JVLA, and the LOw Frequency ARray (LOFAR) radio observations and the {\em XMM-N} X-ray observations. 
The radio characteristics of our sample sources, {\ie}linear extent ($<$40 kpc at $z$ $<$1.2), 
bimodal spectral index (${\alpha}_{\rm 400~MHz}^{\rm 1.5~GHz}$) distribution and the radio luminosities 
(L$_{\rm 1.5~GHz}$ $>$5.0 $\times$ 10$^{23}$ W~Hz$^{-1}$), suggest them to be mainly consist of 
Compact$-$Steep$-$Spectrum (CSS) or Peaked-Spectrum (PS) sources representing an early phase of the AGN-jet activity in dust-obscured environments.  
With stacking we find the existence of faint radio emission 
(S$_{\rm 400~MHz}$ = 72.9~$\mu$Jy~beam$^{-1}$ and S$_{\rm 1.5~GHz}$ = 29~$\mu$Jy~beam$^{-1}$ 
with signal-to-noise ratio $\sim$ 20) in otherwise radio-undetected DOGs. 
Our study revealing the faint emission at a few tens of $\mu$Jy level in high$-z$ DOGs 
can be used as a test-bed for the deeper radio continuum surveys planned with the Square-Kilometer Array (SKA) and its pathfinders.      
\end{abstract}

}]

%
\volnum{43}
\year{2022}
\pgrange{1--17}
\setcounter{page}{1}
\lp{1}
\section{Introduction}
\label{Intro}
Active Galactic Nucleus (AGN), a manifestation of accretion onto the Super-Massive Black Hole (SMBH), is believed to be 
triggered by the inflow of matter into the central region of galaxies. One of the most favored scenarios for AGN triggering 
is galaxy-galaxy interaction or merger that also causes the enhancement in star-formation (SF) activity \citep{Hopkins06}. 
The inter-link between AGN and SF activity is vindicated by the similarity in the cosmological evolution of 
co-moving space-densities of AGN and star-forming galaxies that peak at redshift ($z$) 2$-$3 \citep{Kauffmann2000}. 
It is well established that the intensely star-forming galaxies containing a large amount of gas and dust 
can potentially host obscured AGN (\cite{Lacy20b} and references therein). The large reservoirs of gas and dust absorb 
the optical and UV emission arising from the AGN and stars, and re-radiate it at InfraRed (IR) wavelengths. 
Therefore, the AGN population residing in dusty galaxies can be missed by the optical and 
UV surveys \citep[see][]{Truebenbach17}. The Far-IR (FIR) and Mid-IR (MIR) surveys have been advantageous in 
detecting dust$-$obscured galaxies (DOGs), 
however, the identification of obscured AGN is often challenging. To uncover the obscured AGN, various 
methods, {\eg}hard X-ray detection of AGN emission \citep{Ricci21}, colour selection techniques \citep{Noboriguchi19}, 
MIR emission characterized by a power$-$law \citep{Farrah17}, have been employed. 
Albeit, all these methods suffer from certain limitations; for instance, hard X-ray observations are unable to detect 
AGN with column densities (N$_{\rm H}$) $\geq$10$^{26}$ cm$^{-2}$ due to dominant Compton down-scattering, 
colour selection techniques can be valid up to $z$ $\leq$ 2.0, and 
the MIR spectral energy distributions (SEDs) of intensely star-forming galaxies hosting obscured AGN can be 
fitted equally well with the templates of SFGs as well as of AGN \citep{Shanks21}.       
\par
The {\em Spitzer} and {\em Herschel} observations confirmed that 
DOGs are a subset of optically-faint high-redshift ($z$ $\sim$ 1.5$-$2.5) galaxies with 
their total (8$-$1000~$\mu$m) IR luminosities in the range of 10$^{11.5}$ $L_{\odot}$ $-$ 10$^{14}$ $L_{\odot}$ 
\citep[see][]{Melbourne12}. 
Thus, based on their IR luminosities, DOGs are the most luminous galaxies at their redshifts and can be classified 
as luminous infrared galaxies (LIRGs; $L_{\rm 8-1000~{\mu}m}$ $\geq$ 10$^{11}$~$L_{\odot}$ 
$-$ 10$^{12}$~$L_{\odot}$), ultra-luminous infrared galaxies 
(ULIRGs; $L_{\rm 8-1000~{\mu}m}$ $\geq$ 10$^{12}$~$L_{\odot}$ $-$ 10$^{13}$~$L_{\odot}$), and 
hyper-luminous infrared galaxies 
(HLIRGs; $L_{\rm 8-1000~{\mu}m}$ $\geq$ 10$^{13}$~$L_{\odot}$) 
\citep[see][]{Rowan-Robinson2000,Farrah17}. 
In general, DOGs detected in the {\em Herschel} SPIRE bands, {\ie}250~${\mu}$m, 350~${\mu}$m, and 500~${\mu}$m have 
cooler dust with the temperature of 20$-$40 K than those not detected in the FIR bands. 
To explain the high IR luminosity of DOGs, 
theoretical models invoke gas-rich major merger leading to an intensely star-forming dusty merged system 
\citep{Hopkins08,Yutani22}. The merged system, in the early phase, can be classified as a star-forming DOG, 
while AGN activity can begin once gaseous and dusty material is being fed into the central region hosting SMBH. 
Thus, star-forming DOGs are likely to evolve into AGN-dominated DOGs before eventually turning into quasars or red 
ellipticals \citep[see][]{Dey08}.          
\par 
In the literature, DOGs have been divided into two categories 
{\it viz.} `bump' DOGs (B-DOGs) showing a bump feature centered at 1.6~${\mu}$m in the rest frame, 
and ` power$-$law' DOGs (PL-DOGs) showing 
a power$-$law MIR SED \citep{Melbourne12}. The MIR emission in PL-DOGs is dominated by AGN heated dust, while MIR emission in B-DOGs is mostly powered 
by star-formation \citep{Farrah08}. The obscured AGN are detected more commonly among the extreme population of PL-DOGs characterized by relatively hotter dust emission, also known as the `Hot DOGs' or hyper luminous infrared galaxies 
(L$_{\rm IR}$ $>$ 10$^{13}$ L${\odot}$) residing mostly at high-redshifts ($z$ $\sim$ 2) \citep{Tsai15,Farrah17}. 
However, we cannot rule out the possibility of the existence of obscured AGN in B-DOGs with a less dominant contribution 
to the MIR emission.  
The obscured AGN with radio jets can easily be detected with radio observations as the gas and dust are optically thin at 
radio wavelengths. Therefore, radio observations of DOGs can enable us to unveil hitherto the unexplored population of obscured AGN.      
\par
We point out that the previous radio studies of DOGs have been limited mainly to the radio-powerful PL-DOGs or 
suffer from the shallow flux density limits of radio surveys. 
For instance, \cite{Lonsdale15} performed 
Atacama Large Millimeter/sub-millimeter Array (ALMA) observations at 345~GHz (870 $\mu$m) for 
a sample of 49 dust-obscured quasars selected with ultra-red WISE colour similar to Hot DOGs and compact radio emission in the 1.4 GHz Faint Images of the Radio Sky at Twenty-Centimeters \citep[FIRST;][]{Helfand15} and NRAO VLA Sky Survey \citep[NVSS;][]{Condon98}. 
They found 345 GHz ALMA detection in only 26/49 sources having high star formation rates 
of several thousand M$\odot$ yr$^{-1}$ concurrent with the highly obscured quasar. 
Using 10 GHz Karl G. Jansky Very Large Array (VLA) observations of these radio-powerful dust-obscured quasars,  
\cite{Patil20} reported the evidence for young jets with angular sizes of $<$0.2$^{\prime\prime}$.  
More recently, \cite{Gabanyi21} studied the radio properties of 661 DOGs with the FIRST survey and found only 2$\%$ sources 
of their sample, exclusively PL-DOGs, detected in the FIRST with 5$\sigma$ sensitivity limited to 1.0 mJy beam$^{-1}$. 
The stacking of FIRST image cutouts at the positions of 
the radio-undetected DOGs revealed radio emission only in PL-DOGs at the level of 0.16 mJy beam$^{-1}$ flux density. 
Thus, it is evident that the radio emission in a large fraction of DOGs, even among AGN-dominated DOGs, remained 
undetected due to the relatively shallow detection limit of the FIRST survey. 
With our deep radio observations from the upgraded Giant Metrewave Radio Telescope (uGMRT) we aim to detect radio emission in DOGs at much fainter levels and 
unveil a new population of obscured AGN. 
\par
This paper is structured as follows. In Section~\ref{Data} we provide the details of MIR, optical, and multi-frequency radio data 
used in our study. The sample selection criteria are outlined in Section~\ref{sample}. Radio properties of our sample sources 
are discussed in Section~\ref{Radio}. In Section~\ref{nature} we discuss the nature of our sample DOGs based on the MIR colour-colour diagnostics and X-ray properties. 
In Section~\ref{sec:stack} we demonstrate the existence of faint radio emission in otherwise 
radio-undetected DOGs.   
The importance of much deeper radio surveys from the Square Kilometer Array (SKA) and its 
pathfinders is detailed in Section~\ref{SKA}. The conclusions of our study are given in Section~\ref{Conclusion}.  
\\
In this paper, we adopt cosmological  parameters H$_{0}$ = 70~km~s$^{-1}$~Mpc$^{-1}$, ${\Omega}_{m}$ = 0.3, and 
${\Omega}_{\Lambda}$ = 0.7. Magnitudes are in AB system, unless stated otherwise.  
\section{Infrared, optical and radio data}
\label{Data}
\subsection{IR data}
To select DOGs, we use data from the {\em Spitzer} Wide$-$area InfraRed Extragalactic survey \citep[SWIRE;][]{Lonsdale03} 
and the Hyper Supreme Cam - Subaru Strategic Program  \citep[HSC-SSP;][]{Aihara22} optical survey. 
The SWIRE performed imaging in all four bands of the Infrared Array Camera (IRAC), {\viz}3.6~${\mu}m$, 4.5~${\mu}m$, 
5.8~${\mu}m$ and 8.0~${\mu}m$, and in all three bands of the Multi-band Imaging Photometer (MIPS) {\viz}24~$\mu$m, 70~$\mu$m, 
and 160~$\mu$m. 
The 5$\sigma$ sensitivity of SWIRE is 7.3 $\mu$Jy~beam$^{-1}$, 9.7 $\mu$Jy~beam$^{-1}$, 27.5 $\mu$Jy~beam$^{-1}$ 
and 32.5 $\mu$Jy~beam$^{-1}$ in the four IRAC bands, respectively, 
and 0.45 mJy~beam$^{-1}$, 2.75 mJy~beam$^{-1}$ and 17.5 mJy~beam$^{-1}$ in three MIPS bands, respectively.    
In the {\em XMM-LSS} field, the SWIRE is centered at RA = 02$^{h}$ 21$^{m}$ 00$^{s}$ and 
DEC = -04$^{\circ}$ 30$^{\prime}$ 00$^{\prime}$ (J2000) and covers a total sky$-$area of 8.7 deg$^{2}$ with the IRAC bands 
and 9.0 deg$^{2}$ in the MIPS bands. 
\subsection{Optical data}
In our study, we used optical data from the Hyper Suprime-Cam Subaru Strategic Program 
(HSC-SSP\footnote{https://hsc.mtk.nao.ac.jp/ssp/survey/}) survey, 
which is a multi-band ($g$, $r$, $i$, $z$, $y$ and four narrow-band filters) imaging survey carried out with the 
Hyper Suprime-Cam (HSC) wide-field camera (FOV $\sim$ 1.5~deg$^{2}$) installed on the 8.2-m Subaru telescope \citep{Aihara22}. 
The HSC-SSP survey has three components, {\ie}wide, deep, and ultra-deep, forming a nested coverage wherein the footprints of 
deeper components lie within the footprint of less deep component. 
The wide component covers nearly 1400 deg$^{2}$ around the celestial equator with $i$ band limiting magnitude of 
26.2 mag, while the deep survey component covers a total sky-area of only 27~deg$^{2}$ in 
four different extragalactic fields, including the {\em XMM-LSS} with the limiting magnitude of $m_{\rm i}$ = 26.8 \citep{Aihara22}. 
The ultra-deep component covers a total area of only 3.5~deg$^{2}$ in two sub-fields with the 
limiting magnitude of $m_{\rm i}$ = 27.4. 
We note that the limiting magnitudes are measured with 2$^{\prime\prime}$.0 aperture at 5$\sigma$ level. 
\subsection{Radio data}
{\it 400 MHz band-3 uGMRT observations} : \\
Our band-3 (250 MHz $-$ 550 MHz) continuum imaging observations with the uGMRT were carried out 
during 11 September 2017 
under the proposal code 32$\_$066.
These observations of one pointing centered at RA = 02$^{h}$ 26$^{m}$ 45$^{s}$.0 and 
DEC = -04$^{\circ}$ 41$^{\prime}$ 30$^{\prime\prime}$.0 were performed in the full synthesis mode with a total observing time 
of 10 hours. The flux calibrators 3C48 and 3C147 were observed for 15$-$20 minutes at the beginning and end of the observing 
session, respectively. 
The phase calibrators 0116-0208 and 0323+055 were observed for 5$-$6 minutes with each scan inter$-$leaved with 30 minutes 
scan of the target field.  
The data were reduced using a Common Astronomy Software Applications (CASA\footnote{https://casa.nrao.edu/}) based pipeline. 
The standard data reduction procedure includes flagging of bad data, calibration of visibilities, and imaging 
of calibrated-visibilities. 
The final image was obtained after applying multiple iterations of `phase-only' and `phase-and-amplitude' self$-$calibration 
meant for correcting artifacts arising from the residual calibration errors.
The primary beam correction was applied to increase the fidelity of the image in the outer regions.  
Our final image has a median noise-rms of 30 $\mu$Jy beam$^{-1}$, although noise-rms around 
relatively bright sources ($>$100 mJy) and at the peripheral regions is higher. 
The final image has synthesized beam of 6$^{\prime\prime}$.7 $\times$ 5$^{\prime\prime}$.3 with position angle of 71$^{\circ}$.7, 
and covers a total sky area of nearly 2.4 deg$^{2}$ consisting a deeper central region of 1.3 deg$^{2}$ with 
noise-rms $\leq$40 $\mu$Jy beam$^{-1}$. 
\par
We created the source catalogue by using the Python Blob Detector and Source Finder \citep[PyBDSF;][]{Mohan15} algorithm. 
While running the PyBDSF we set the source detection limit to 5$\sigma$ for peak pixel and 3$\sigma$ for island boundary 
and grouped overlapping Gaussians into a single source. 
To minimize the number of spurious sources around bright sources we adopt smaller rms box of 33 $\times$ 33 pixel$^{2}$ 
with sliding of 11 pixels in the regions containing bright sources, while rms box size of 40 $\times$ 40 pixel$^{2}$ with sliding 
of 13 pixels was used in the remaining part of the image.       
With the aforementioned parameters we obtain a catalogue of 2332 sources detected at $\geq$5$\sigma$ level. 
\\ \\
{\it Ancillary radio data} : For our study, we also used the radio data at other frequencies. 
The footprints of uGMRT observations are covered with the 1.5 GHz Jansky Very Large Array (JVLA) 
and 150 MHz LOw Frequency ARray (LOFAR) observations available in the {\em XMM-LSS} field. 
The 1.5 GHz wide-band (0.994$-$2.018 GHz) JVLA 
radio observations achieved a median noise-rms of 16 $\mu$Jy beam$^{-1}$ and 
angular resolution of 4$^{\prime\prime}$.5 \citep{Heywood20}. 
The 150 MHz LOFAR observations reported by \cite{Hale19} cover the sky-area of 27~deg$^{2}$ with 
mosaiced image having a median noise-rms in the range of 0.28~mJy~beam$^{-1}$ to 0.40 mJy~beam$^{-1}$ 
and the synthesized beam of 7$^{\prime\prime}$.5 $\times$ 8$^{\prime\prime}$.5. 
We note that the sky-area of our uGMRT observations is also covered by the shallow large-area surveys such as 1.4 GHz FIRST and 3.0 GHz Very Large Area Sky Survey 
(VLASS). The FIRST performed with VLA B-configuration has the resolution of 5$^{\prime\prime}$.0 and noise-rms 0.15 mJy~beam$^{-1}$ 
\citep{Becker95}. The more recent VLASS is a high-frequency analogue of the FIRST and provides 
a median noise-rms of 0.12 mJy~beam$^{-1}$ and angular resolution of 2$^{\prime\prime}$.5 \citep{Lacy20}. 

\section{Dust-obscured galaxies : Selection criteria, sample and redshift estimates}
\label{sample}
\subsection{Selection criteria}
In the literature, DOGs have been identified by using the flux ratio of FIR to optical \citep{Toba15}. 
Using {\em Spitzer} survey data \cite{Fiore08} demonstrated that the flux ratio of 24~$\mu$m to $R$ band optical 
(S$_{\rm 24~{\mu}m}$/S$_{\rm R}$) $\geq$1000 can efficiently select DOGs. 
The cutoff limit of 1000 is based on the fact that it selects sources redder 
than the colour of most ULIRGs at all redshifts \citep{Dey08}.   
The flux ratio criterion corresponds to the colour cut of $R$ - [24] $\geq$7.5; where magnitudes are measured in the AB system. 
\begin{figure*}[ht]
\includegraphics[angle=0,width=8.0cm,trim={0.0cm 6.5cm 0.0cm 7.0cm},clip]{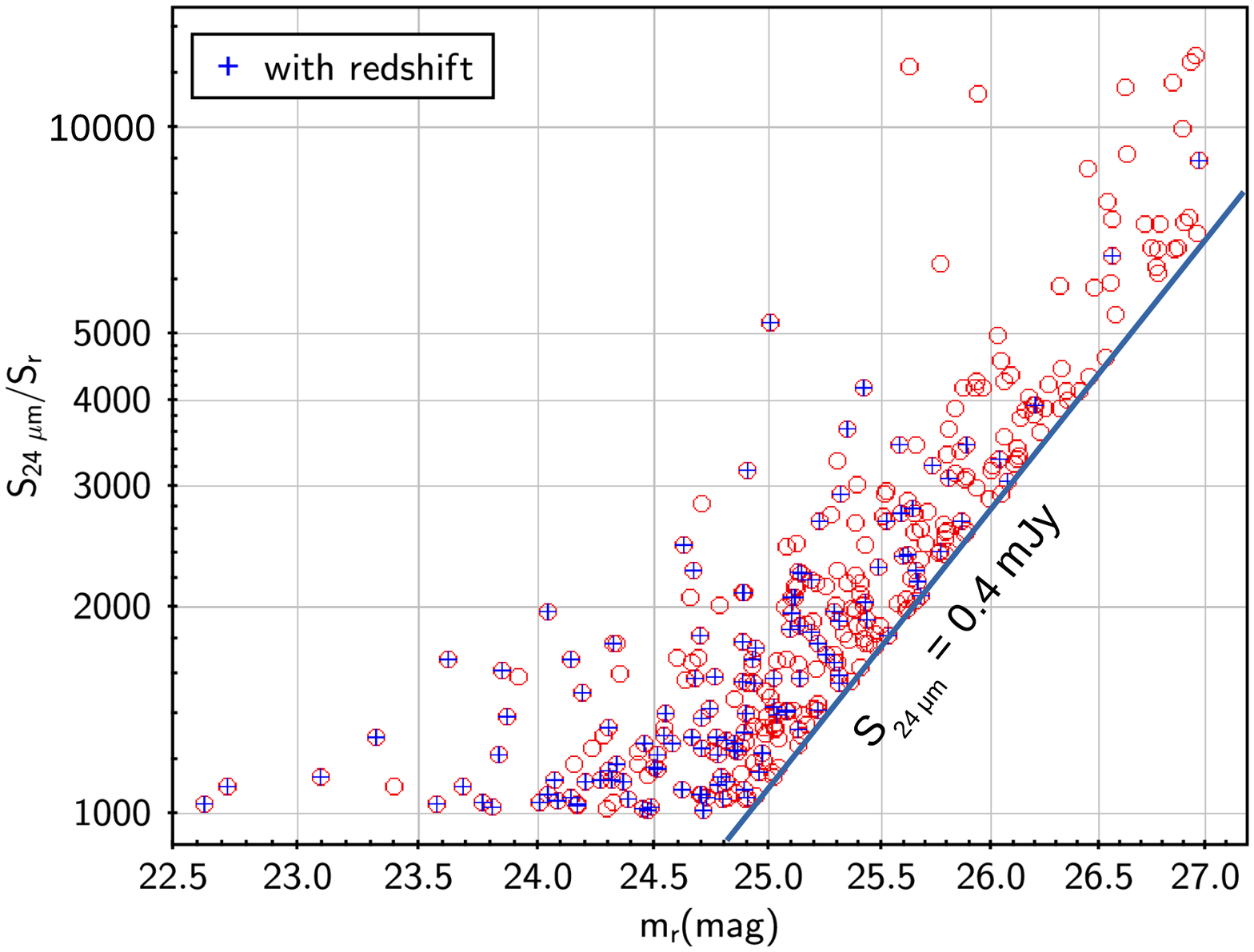}
{\includegraphics[angle=0,width=8.0cm,trim={0.0cm 0.0cm 0.0cm 0.0cm},clip]{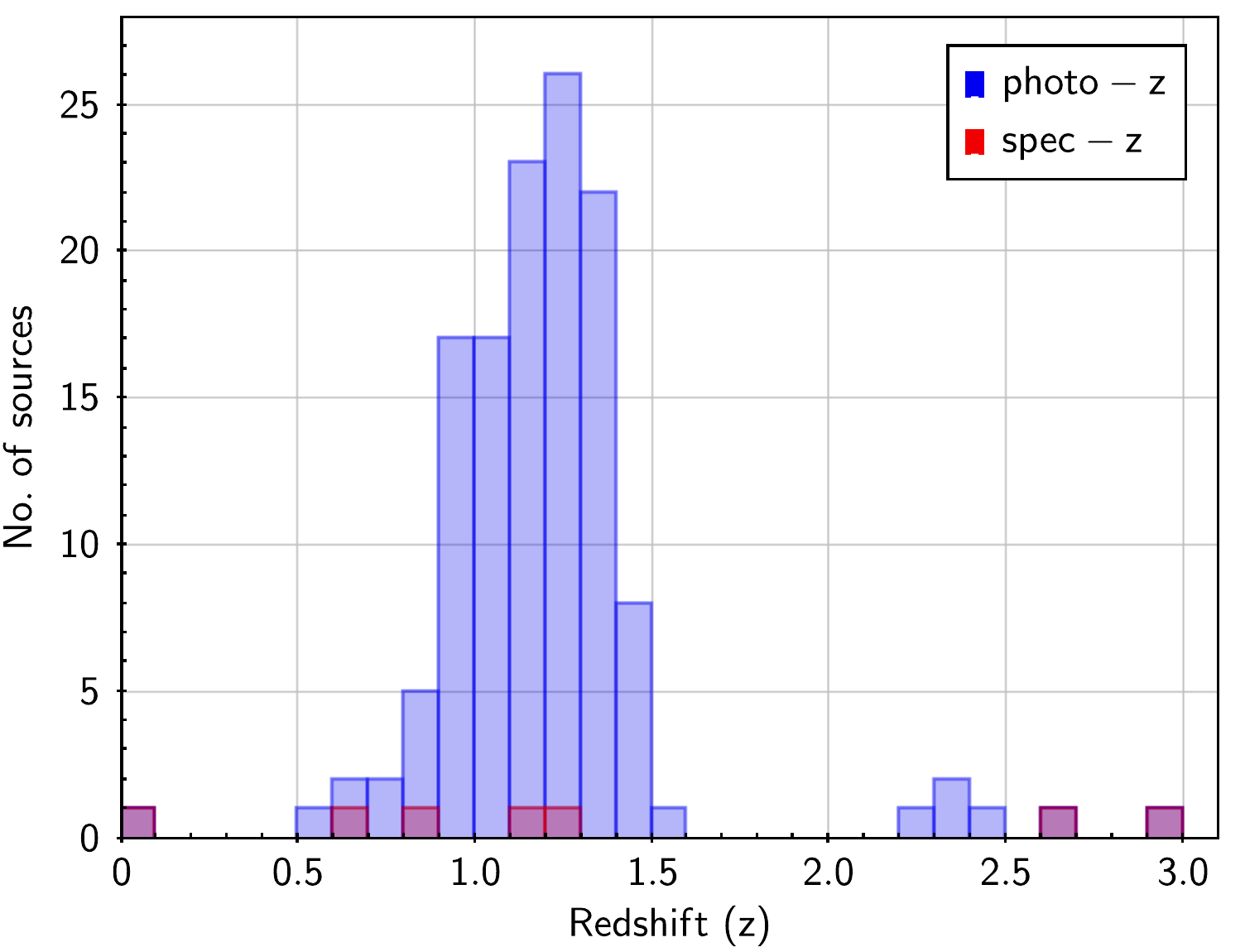}}
\caption{{\it Left panel}: The flux ratio of 24~${\mu}$m to $r$ band (S$_{\rm 24~{\mu}m}$/S$_{r}$) 
versus $r$ band magnitude ($m_{r}$). {\it Right panel}: The histogram of available redshift estimates for our sample DOGs.}
\label{Redshift}
\end{figure*}
To select DOGs, we first identified the optical counterparts of 24~$\mu$m sources by cross-matching the 
24~$\mu$m SWIRE catalogue with the HSC-SSP optical catalogue. 
We use IRAC 3.6~$\mu$m counterparts positions of 24~$\mu$m sources, whenever available, considering the fact that 
the 3.6~$\mu$m IRAC observations provide better angular resolution (2$^{\prime\prime}$.0) than that of 
the MIPS at 24~$\mu$m (5$^{\prime\prime}$.6). 
Thus, the optical counterpart is searched around the 3.6~$\mu$m source position within a radius of 
1$^{\prime\prime}$.0 and the closest source is being considered as the true counterpart. 
In case of the unavailability of IRAC coverage we search optical counterparts around the 24~$\mu$m source position within 
a search radius of 2$^{\prime\prime}$.0. In our cross-matching exercise, we followed the method described in 
\cite{Singh15} to choose the optimum search radius which in turn ensures maximum completeness with minimum contamination by random matches. 
%
After identifying the optical counterparts of 24~$\mu$m sources, 
we derive a sample of 321 DOGs that satisfy the following criteria : 
(i) S$_{\rm 24~{\mu}m}$/S$_{r}$ $\geq$1000, where S$_{\rm 24~{\mu}m}$ and S$_{r}$ are 24~$\mu$m and $r$ band 
fluxes, respectively; (ii) S$_{\rm 24~{\mu}m}$ $\geq$0.4 mJy (24~$\mu$m sample suffers incompleteness below 
this flux limit); (iii) sources fall within the region covered by our 400 MHz uGMRT observations. 
Table~\ref{Radiosample} shows a representative sub-sample of our DOGs detected at radio wavelengths. 
\par
\begin{table*}[ht]
\caption{A representative sub-sample of 10 radio-detected DOGs.}
\label{Radiosample} 
\scalebox{0.72}{
\begin{tabular}{lcccccccccc}
\topline
RA  & DEC & S$_{\rm 24~{\mu}m}$ & $m_{\rm r}$ & S$_{\rm 24~{\mu}m}$/S$_{r}$ & S$_{\rm 400~MHz}$ & S$_{\rm 1.5~GHz}$ & 
${\alpha}_{\rm 400~MHz}^{\rm 1.5~GHz}$ & $z$   &  logL$_{\rm 400~MHz}$ & logL$_{\rm 1.5~GHz}$ \\ 
(h m s) & (d m s)  &  (mJy)    &   (mag)     &       &   (mJy)    &  (mJy)   &        &    &  (W~Hz$^{-1}$)  & (W~Hz$^{-1}$) \\
\hline
02:23:23 & -04:49:26 & 0.519$\pm$0.022& 24.93$\pm$0.03 & 1342.3 & 1.37$\pm$0.17 & 0.383$\pm$0.006 & -0.96$\pm$-0.22 & $>$1.0 & $>$24.84 & $>$24.29\\
02:23:24 & -04:26:03 & 0.550$\pm$0.025 & 24.81$\pm$0.03 & 1273.3 & 0.75$\pm$0.15 & 0.292$\pm$0.006 & -0.71$\pm$-0.36 & 1.17$\pm$0.14 (p) & 24.66 & 24.26 \\
02:23:42 & -04:43:56 & 0.797$\pm$0.024 & 24.63$\pm$0.04 & 1565.3 & 0.65$\pm$0.20 & 0.176$\pm$0.004 & -0.99$\pm$-0.54 & $>$1.0  & $>$24.53 & $>$23.96\\
02:23:47 & -04:20:46 & 0.921$\pm$0.025 & 24.30$\pm$0.03 & 1335.6 & 0.57$\pm$0.15 & 0.099$\pm$0.005 & -1.33$\pm$-0.48 & 
1.25$\pm$0.13 (p) & 24.84 & 24.08 \\
02:23:49 & -05:04:53 & 0.937$\pm$0.022 & 24.01$\pm$0.02 & 1034.3 & 0.89$\pm$0.14 & 0.349$\pm$0.006 & -0.71$\pm$-0.28 & 0.0336$\pm$0.0001 (s) & 21.36 & 20.96\\
02:23:54 & -04:14:37 & 1.340$\pm$0.027 & 23.68$\pm$0.01 & 1096.2 & 0.57$\pm$0.17 & 0.131$\pm$0.004 & -1.11$\pm$-0.54 & 1.32$\pm$0.12 (p) & 24.81 & 24.17 \\
02:24:24 & -05:08:37 & 0.559$\pm$0.025 & 25.85$\pm$0.15 & 3381.9 & 0.57$\pm$0.15 & 0.150$\pm$0.004 & -1.01$\pm$-0.46 & $>$1.0  & $>$24.48 & $>$23.89 \\
02:24:26 & -04:48:29 & 0.602$\pm$0.025 & 25.13$\pm$0.06 & 1878.2 & 1.06$\pm$0.09 & 0.328$\pm$0.005 & -0.89$\pm$-0.16 & 1.12$\pm$0.12 (p) & 24.83 & 24.32 \\
02:24:28 & -04:33:53 & 0.635$\pm$0.029 & 26.15$\pm$0.17 & 4584.3 & 0.52$\pm$ 0.11 & 0.150$\pm$0.005 & -0.94$\pm$-0.37 & $>$1.0 & $>$24.41 & $>$23.88 \\
02:24:35 & -05:11:27 & 0.633$\pm$0.023 & 26.32$\pm$0.12 & 5879.1 & 0.79$\pm$0.09 & 0.205$\pm$0.004 & -1.02$\pm$-0.22 & $>$1.0 & $>$24.62 & $>$24.04 \\             
\hline
\end{tabular}}
\tablenotes{{\it Notes} - Only first ten sources of the radio-detected DOGs are listed in the increasing order of RA. 
The spectroscopic and photometric redshifts are indicated, within brackets, by `s' and `p', respectively.} 
\end{table*}
We note that the DOGs with no optical ($r$ band) counterparts can be extremely obscured galaxies, although, 
we find that all our sample sources show optical counterparts in the HSC-SSP. We excluded sources falling within the masked regions in the HSC-SSP optical image.  
From S$_{\rm 24~{\mu}m}$/S$_{r}$ versus $r$ band magnitude ($m_{r}$) plot (see Figure~\ref{Redshift}, left panel) it is evident 
that the DOGs in our sample are generally faint in the optical with $r$ band magnitude distributed in the range 
of 22.63 to 26.97 with a median value of 25.20. We also find that the faint optical counterparts have apparently higher values of S$_{\rm 24~{\mu}m}$/S$_{r}$ due to 
the 24~$\mu$m flux cutoff limit applied in our sample selection criteria. 
We find that 24~$\mu$m luminosities for our sample sources, with available redshifts, are found to be in 
the range of 7.96 $\times$ 10$^{7}$ L$\odot$ to 2.35 $\times$ 10$^{12}$ L$\odot$ with a median value 
of 1.56 $\times$ 10$^{11}$ L$\odot$. All but one of our sources with available redshifts have 
24~$\mu$m luminosity higher than 4.8 $\times$ 10$^{10}$ L$\odot$ (see Table~\ref{Radiosample}). 
Therefore, based on the total IR luminosities, our sources can be classified as LIRGs and ULIRGs. 
\subsection{Redshift estimates}
We checked the availability of redshifts for our DOGs using spectroscopic and photometric redshift catalogues given in 
the HSC-SSP data access site\footnote{https://hsc.mtk.nao.ac.jp/ssp/data-release/}.
The HSC$-$SSP PDR3 provides spectroscopic redshifts for the HSC-SSP sources, whenever available (spec$-z$ table). 
The spectroscopic redshifts are gleaned from the publicly available spectroscopic redshifts surveys such as the VIMOS$-$VLT Deep Survey \citep[VVDS;][]{LeFevre13}, 
PRIsm MUlti$-$object Survey (PRIMUS) DR1 \citep{Cool13}, VIMOS Public Extragalactic Redshift Survey \citep[VIPERS;][]{Garilli14} in the {\em XMM$-$LSS} field.
We also used the HSC-SSP PDR2 photometric redshift (photo$-z$) catalogue that provides redshifts obtained by using ﬁve-band HSC 
photometry. The photometric redshifts are derived from a photo$-z$ code \citep[MIZUKI;][]{Tanaka18} based on 
the SED ﬁtting technique. 
To avoid highly inaccurate photo$-z$ estimates we used photo$-z$ estimates with reduced ${\chi}^{2}$ $\leq$1.5 in 
the template ﬁtting, and the fractional error on the derived redshift ${\sigma}_{z}$/$z$ $\leq$ 0.25. 
For our sample sources with no available redshifts in the photo$-z$ catalogue, 
we used the photo$-z$ catalogue provided by \citep{Schuldt21}, who derived redshift by using a Convolutional 
Neural Network (CNN) 
method that uses galaxy images in addition to five band HSC-SSP photometric points. The CNN based photo$-z$ estimates have an accuracy of $|z_{\rm pred} - z_{\rm ref}|$ = 0.12 for the full HSC$-$SSP photometric in the redshift range of 0 to 4.0. 
\par
We attempt to find redshift estimates of our sample sources by cross-matching them with the available 
spectroscopic and photometric redshift catalogues. We considered the closest match within 1$^{\prime\prime}$.0 
radius centered at the optical positions of our sample sources. 
The cross-matching yielded spectroscopic redshifts for only seven sources that are derived from the 
VVDS (for five sources) and the PRIMUS (for two sources), while photometric redshifts are found for 124 sources. 
Thus, redshift estimates are available for only 131/321 (40.8$\%$) of our DOGs. 
Figure~\ref{Redshift} ({\it right panel}) shows the redshift distribution 
spanning in the range 0.03 to 2.99 with a median value of 1.19. We find that 
the redshifts are mostly limited up to 1.5, and only seven sources have redshifts higher than 1.5. 
The lack of sources at higher redshift may be attributed to the larger inaccuracy in photo-$z$ 
estimates and the faintness in the optical band for high$-z$ sources. 
The comparison of $r$ band magnitudes shows that the sources with redshift estimates have $m_{r}$ in the range 
of 22.63 to 26.97 with a median value of 24.88, while sources with no redshifts are systematically fainter in 
the optical bands with $m_{r}$ lying in the range of 23.39 to 26.96 with a median value of 25.49.
Also, for high$-z$ sources (z $>$2.0) optical bands would sample the rest-frame UV light, 
which is strongly obscured by the dust present in host galaxies. Thus, in comparison to their 
low$-z$ counterparts, the high$-z$ DOGs are expected to be even much fainter in the optical.
\section{Radio characteristics of DOGs}
To investigate the radio emission characteristics of DOGs, we identified the radio counterparts of DOGs using 400 MHz uGMRT, 1.5 GHz JVLA, and 150 MHz LOFAR observations. 
\label{Radio}
\subsection{Radio detection rates}
\label{sec:RadioDet}
We cross-matched our sample sources with the 400 MHz uGMRT, 1.5 GHz JVLA and 150 MHz LOFAR source catalogues using 
an optimum search radius of 2$^{\prime\prime}$.0 to 2$^{\prime\prime}$.5. To ensure the reliable cross-matching we 
also visually inspected the uGMRT and JVLA radio images of our DOGs counterparts.  
\begin{figure*} [ht]
\includegraphics[angle=0,width=17.5cm,trim={0.0cm 0.0cm 0.0cm 0.0cm},clip]{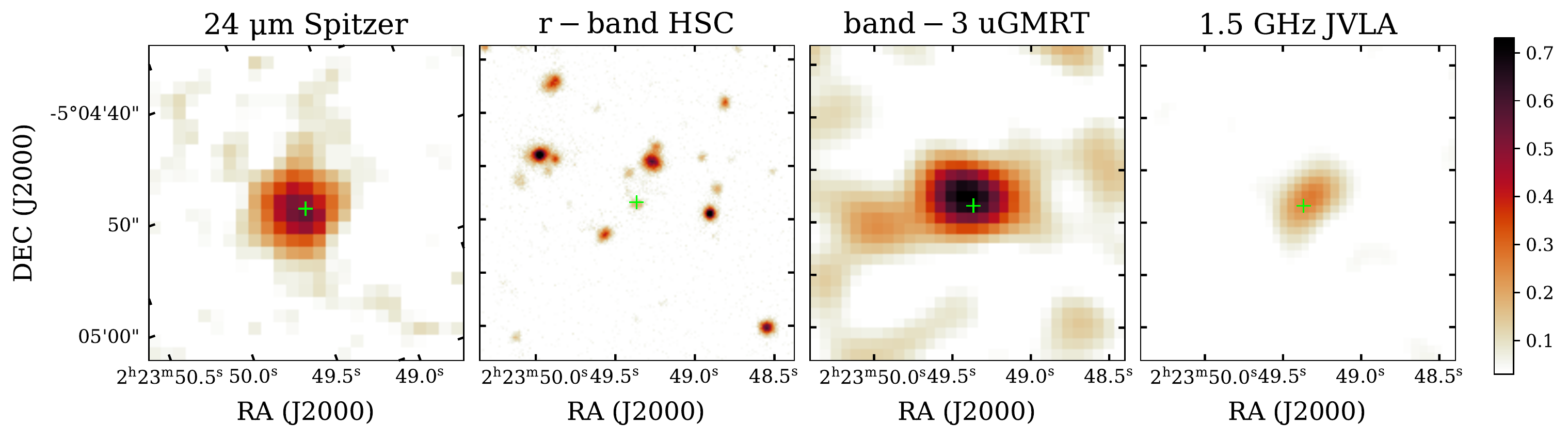}
\includegraphics[angle=0,width=17.5cm,trim={0.0cm 0.0cm 0.0cm 0.0cm},clip]{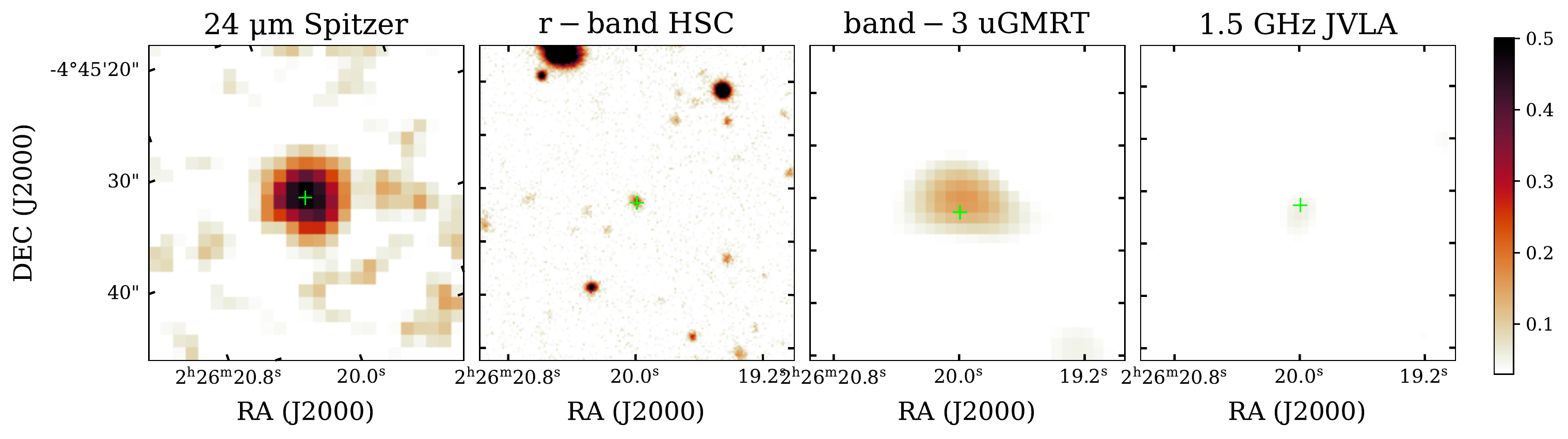}
\caption{The 24~$\mu$m FIR, $r$ band optical, 400 MHz uGMRT and 1.5 GHz JVLA radio images of 
two representative DOGs from our sample. The upper panel shows a case where source is detected in both 400 MHz uGMRT 
(S$_{\rm 400~MHz}$ = 0.89$\pm$0.14 mJy) and 1.5 GHz JVLA (S$_{\rm 1.5~GHz}$ = 0.35$\pm$0.01 mJy). 
The lower panel presents a case where source is detected only in the uGMRT (S$_{\rm 400~MHz}$ = 0.22$\pm$0.04 mJy) 
but falls below the 5$\sigma$ detection limit of the JVLA, although a faint emission is apparent. 
All the images are of 30$^{\prime\prime}$ $\times$ 30$^{\prime\prime}$. The plus symbol represents the optical position of sources. A vertical colour bar indicates intensity level in mJy beam$^{-1}$}.
\label{fig:MWImages}
\end{figure*}
\begin{table*}[ht]
\caption{The radio detection rates in our sample of 321 DOGs.}
\label{RadioDet} 
\begin{tabular}{lcccccc}
\topline
Radio        & Frequency   &   BW      &  5$\sigma$ depth & Resolution & N$_{\rm Detected}$ &  Detection rate  \\  
observations &  (MHz)      &  (MHz)    & (mJy~beam$^{-1}$) & ($^{\prime\prime}$$\times$$^{\prime\prime}$)  &    & (per cent) \\ \midline
uGMRT        &  400        &   200     &  0.15        & 6.7$\times$5.3 & 91/321  & 28.3  \\
JVLA         &  1500       &   1000    &  0.08        & 4.5$\times$4.5 & 83/321  & 25.8  \\
LOFAR        &  150        &   130     &  1.4$-$1.97  & 7.5$\times$8.5 & 11/316  & 3.5   \\
FIRST        &  1400       &   128     &  1.0         & 5.0$\times$5.0 & 02/321  & 0.6  \\ 
VLASS        &  3000       &   2000    &  0.6         & 2.5$\times$2.5 & 03/321  & 0.9  \\ 
All (uGMRT or JVLA) &      &           &              &                & 109/321 & 34     \\
\hline
\end{tabular}
\end{table*}
We find that only 91/321 (28.3 per cent) DOGs are detected in the 400~MHz uGMRT observations.  
The detection rate in the 1.5 GHz JVLA observations is a little lower, with only 83/321 (25.8 per cent) 
detected sources. 
Notably, only 65 sources are detected in both uGMRT and JVLA observations, 
while 26 sources are detected in the uGMRT observations but not in the JVLA observations, and 18 sources are detected in the 
JVLA observations but not in the uGMRT observations. 
Thus, the radio detection rate of our sample DOGs is 109/321 $\simeq$34 per cent if detection 
at one frequency (either 400 MHz or 1.5 GHz) is considered. 
We note that our detections are limited to $\geq$5$\sigma$ level, and hence, 
non$-$detection can still show a faint emission falling below the 5$\sigma$ limit. 
In Figure~\ref{fig:MWImages} we show 24~$\mu$m MIPS, $r$ band HSC-SSP, 400~MHz uGMRT 
and 1.5 GHz JVLA images of our two sample sources. 
In J022349-050453, we detect radio counterparts in both 400~MHz uGMRT 
and 1.5 GHz JVLA observations (see Figure~\ref{fig:MWImages} {\it upper panel}). 
While, in J022619-044536, we detect only a faint emission at 1.5 GHz falling below the 5$\sigma$ 
detection limit (see Figure~\ref{fig:MWImages} {\it bottom panel}). 
\par
In Table~{\ref{RadioDet}} we list the basic parameters (frequency, bandwidth, sensitivity 
and resolution) of different radio observations and the detection rates of our sample DOGs in them. 
We note that a somewhat higher detection rate in the 400~MHz uGMRT observations than that in the 1.5 GHz JVLA can be attributed 
to its better sensitivity. The 5$\sigma$ median sensitivity of 0.15 mJy~beam$^{-1}$ of our 400~MHz
uGMRT observations correspond to 0.06 mJy~beam$^{-1}$ at 1.5 GHz, assuming a typical spectral index of -0.7. 
Thus, our 400~MHz uGMRT observations are deeper than 1.5 GHz JVLA observations having an average 5$\sigma$ depth 0.08~mJy~beam$^{-1}$.  
Moreover, the detection of several sources at only one frequency 
(in 400~MHz uGMRT but not in 1.5~GHz JVLA or vice versa) can be due to the diverse 
spectral properties of radio sources, {\ie}steep sources are preferentially detected in 
the uGMRT observations while JVLA observations can detect flat spectrum sources 
(see Section~\ref{sec:spectra} for more discussion). 
\par
We note that (316/321) majority of our sample sources fall within the 150 MHz LOFAR footprints; however, only 
11/316 (3.5 per cent) sources show detection in the LOFAR observations. 
All the sources detected in the LOFAR observations are relatively bright 
(S$_{\rm 150~MHz}$ $>$1.5 mJy) and are also detected in both 400~MHz uGMRT as well as 1.5 GHz 
JVLA observations.  
A much lower detection rate in the LOFAR observations is due to its relatively lower sensitivity and 
the faintness of radio emission in our sample DOGs.  
We also checked the radio counterparts of our sample sources in the 1.4 GHz FIRST and 3 GHz VLASS, and 
found only two and three sources, respectively.  
Thus, the detection rate of our sample sources in the FIRST and VLASS is $<$1.0 per cent. 
We point out that our results are consistent with the findings of \cite{Gabanyi21} who reported that only 2 per cent 
of their {\em WISE}-selected IR-bright DOGs are detected in the FIRST. 
For our sample sources, the lower detection rate ($<$1.0 per cent) found in the FIRST can be because 
that our {\em Spitzer}-selected DOGs are much fainter in the FIR than the {\em WISE}-selected DOGs. 
This is further vindicated by the fact that, in our sample itself, we find that the radio-detected DOGs are somewhat brighter at 24~${\mu}$m than those 
not detected in the radio. In our sample, the radio-detected DOGs have 24 $\mu$m fluxes in the range 
of 0.41 mJy to 3.2 mJy with a median value of 0.6 mJy, while, the radio undetected DOGs 
have 24 $\mu$m fluxes distributed over a similar range (0.4 mJy to 3.3 mJy) but 
have a lower median value of 0.52 mJy. 
%
\subsection{Flux densities}
We examine the flux density distributions of our sample sources in different radio observations. 
\begin{figure} [ht]
\includegraphics[angle=0,width=8.0cm,trim={0.0cm 0.0cm 0.0cm 0.0cm},clip]{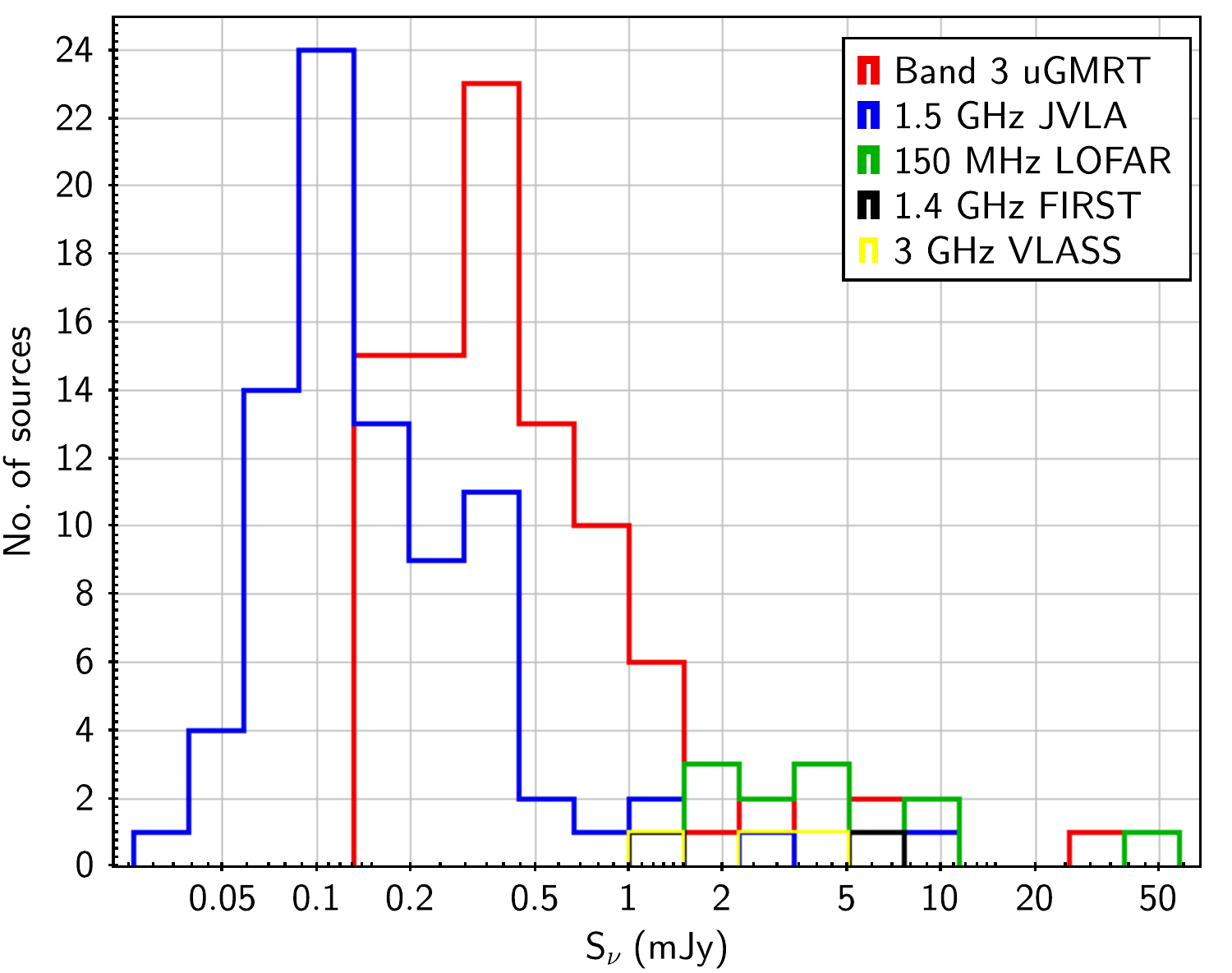}
\caption{The flux density distributions of radio-detected sources in different observations.}
\label{fig:FD}
\end{figure}
Figure~\ref{fig:FD} shows the flux density distributions of our radio-detected DOGs at 400 MHz, 1.5 GHz and 150 MHz. 
Table~\ref{tab:range} lists the range and median values of flux density distributions.
We find that 400 MHz flux densities of 91/321 radio-detected DOGs are distributed in the range of 0.137 mJy to 26.88 mJy with 
a median value of 0.37 mJy. The 1.5 GHz flux densities are found in the range of 0.035 mJy to 9.33 mJy, with a median value 
of 0.13 mJy. Unlike uGMRT and JVLA observations, the LOFAR detections are limited only to relatively bright 11 sources with 150 MHz 
flux densities in the range of 1.56 mJy to 40 mJy with a median value of 3.75 mJy. 
We note that the flux density distributions at different frequencies may not offer a direct comparison, 
although both 400 MHz and 1.5 GHz flux density distributions show that the radio emission in our sample DOGs are generally faint at the level of sub-mJy that remained mostly undetected in the FIRST and VLASS surveys. 
\begin{table}[ht]
\caption{The range and median values of various parameters.}
\label{tab:range} 
\scalebox{0.85}{
\begin{tabular}{lccc}
\topline
Parameter (unit)                       & N$_{\rm sources}$  &  Range      & Median   \\ \midline  
\multicolumn{3}{c}{Optical properties}                                                  \\
    $m_{r}$                            &      321        &  22.63$-$26.97 & 25.21  \\
     redshift ($z$)                    &      131        &  0.034$-$2.99  & 1.19   \\ 
\multicolumn{3}{c}{IR properties}                                                  \\
S$_{\rm 24~{\mu}m}$ (mJy)              &      321        &  0.4$-$3.3  &  0.55    \\
log($\nu$L$_{\rm 24~{\mu}m}$) (L$\odot$)&      131        &  7.9$-$12.37   & 11.19   \\   
\multicolumn{3}{c}{Radio properties}                                           \\
S$_{\rm 400~MHz~uGMRT}$ (mJy) &       91        &  0.137$-$26.88 & 0.37   \\
S$_{\rm 1.5~GHz,~JVLA}$ (mJy) &       83        &  0.035$-$9.33  & 0.13    \\
S$_{\rm 150~MHz,~LOFAR}$ (mJy) &       11        &  1.56$-$40.0   & 3.75    \\
${\alpha}_{\rm 400~MHz}^{\rm 1.5~GHz}$ &       65        & -1.89$-$0.02   & -0.91  \\
log(L$_{\rm 400~MHz}$) (W~Hz$^{-1}$)   &       44        &  21.36$-$25.98 & 24.45  \\            
log(L$_{\rm 1.5~GHz}$) (W~Hz$^{-1}$)   &       39        & 20.95$-$25.52  & 23.95  \\ 
\multicolumn{3}{c}{X-ray properties}                                                  \\
log(S$_{\rm 0.5-10~keV}$)             & 24   &  -14.45$-$-13.43 & -13.97  \\ 
log(L$_{\rm 0.5-10~keV}$) (erg~s$^{-1}$)&       24        &  43.36$-$45.51 & 44.31  \\     
HR$_{\rm 0.5-2.0~keV}^{\rm 2.0-10~keV}$&       24        & -0.11$-$0.91   & 0.76  \\         
\hline
\end{tabular}}
\tablenotes{Notes - The X-ray flux is measured in the unit of erg~cm$^{-2}$~s$^{-1}$.} 
\end{table}
It is worth mentioning that using stacking of FIRST image cutouts, \cite{Gabanyi21} reported 
the median 1.4 
flux density to be nearly 0.16 mJy in their sample of IR-bright PL-DOGs. 
Our 1.5 GHz JVLA observations detect all the sources with flux density above 0.16 mJy.  
Hence, we emphasize that our deep uGMRT and JVLA observations allow the direct detection of radio emission in DOGs 
that was only anticipated from the stacking of the FIRST images.   
\subsection{Radio sizes and morphologies}
\label{sec:size}
The size and morphology of radio emission are important parameters for unveiling the nature of a radio source. 
A typical double-lobe radio morphology can be considered as an unambiguous signature of AGN-jet activity. 
The radio size of AGN-jet system can also infer the evolutionary stage of radio source \citep{An12}. 
For our sample sources, the source extraction algorithm provides the source size measurements by fitting one or more two-dimensional 
elliptical Gaussians to the radio emission.
\begin{figure*}[ht]
\includegraphics[angle=0,width=8.0cm,trim={0.0cm 0.0cm 0.0cm 0.0cm},clip]{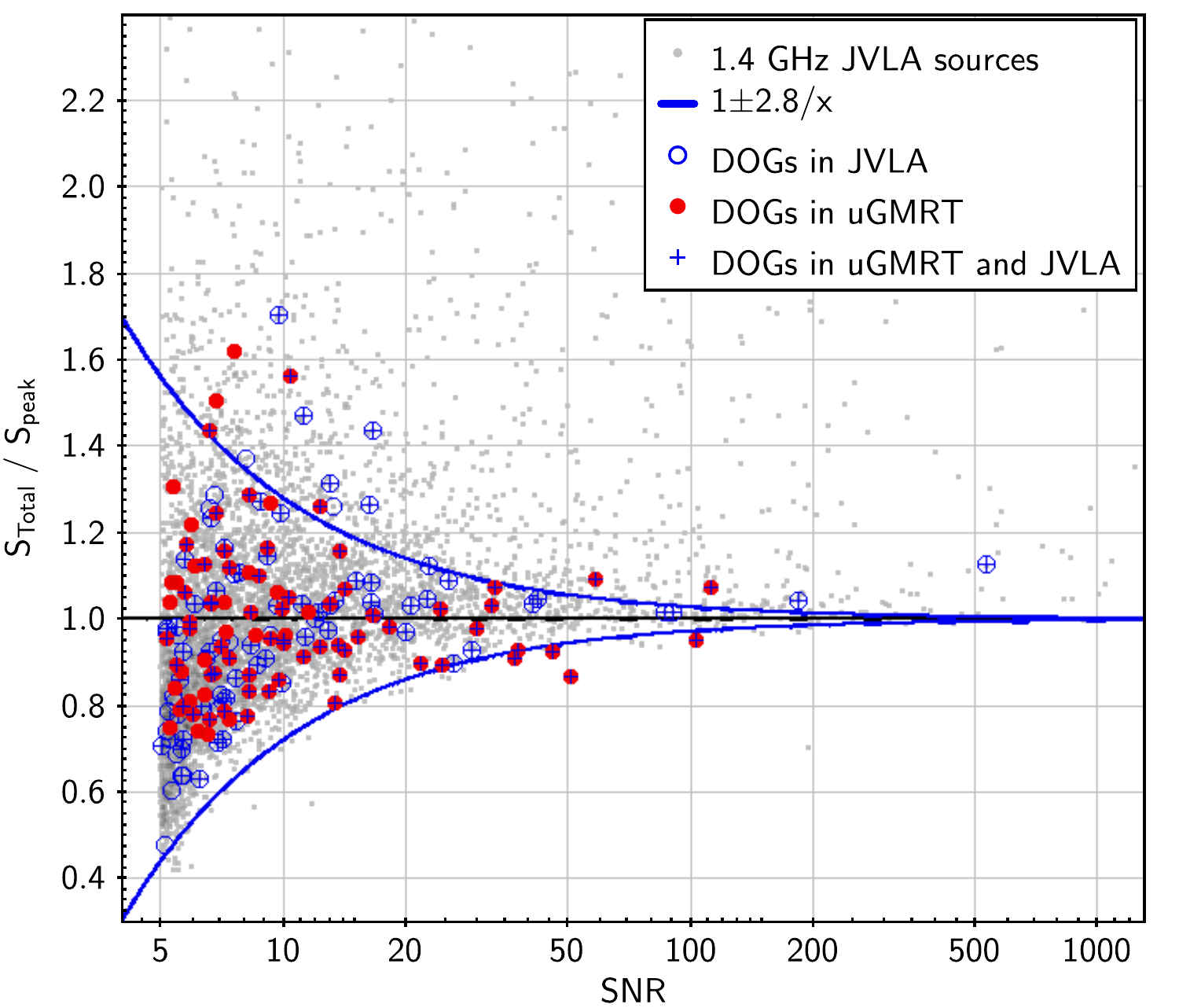}
{\includegraphics[angle=0,width=9.0cm,trim={0.0cm 6.5cm 0.0cm 8.0cm},clip]{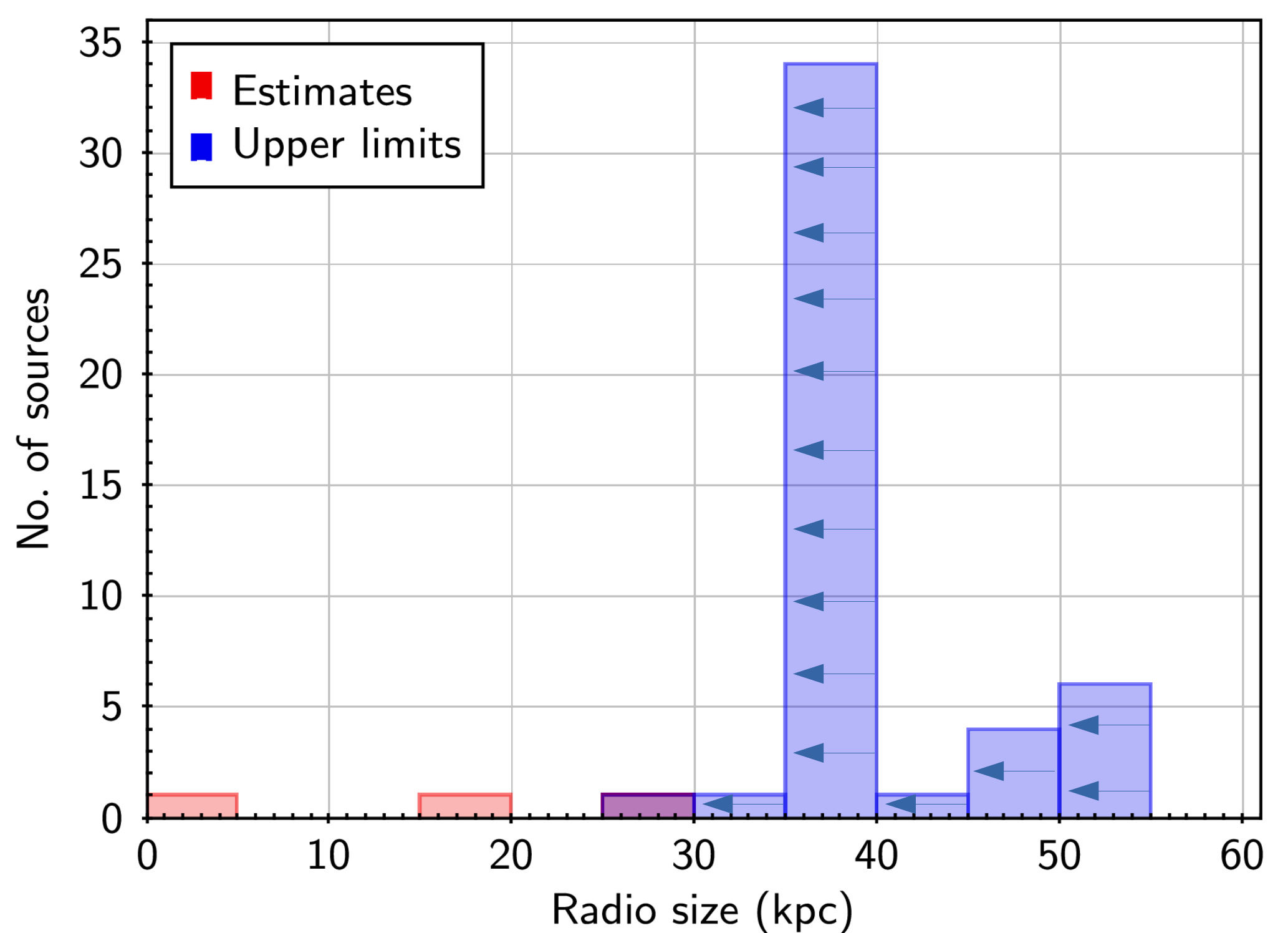}}
\caption{{\it Left panel}: The ratio of total-to-peak flux density versus SNR plot. Sources falling within the envelope 
of blue curves are unresolved point sources. The sources detected in both uGMRT and JVLA observations appear twice as the 
values based on both observations are shown. An offset on the ratio of total-to-peak flux density seen in the uGMRT 
observations is adjusted to 1. The y-axis is restricted to 2.4 for highlighting the trend shown by the faint point sources. 
{\it Right panel} : The histogram of physical radio sizes}.
\label{size}
\end{figure*}
However, the presence of noise peaks around a source can affect the fitted size, particularly in the case of faint sources. To account for the errors introduced by noise peaks, we use the diagnostic plot 
of the total-to-peak flux density ratio versus signal-to-noise ratio (SNR) (see Figure~\ref{size} ({\it Left panel})). This method is 
commonly used in the studies involving a large faint radio source population \citep[see][]{Singh18,Ishwar20}.   
We note that the unresolved point sources are expected to have S$_{\rm total}$/S$_{\rm peak}$ = 1.0. 
Although, 
Figure~\ref{size} ({\it Left panel}) shows an increasingly large dispersion around  S$_{\rm total}$/S$_{\rm peak}$ = 1.0 line for 
fainter sources that 
suffer higher fractional error caused by the negative and positive noise peaks present around them. 
To account for the errors, we plot a curve defined as $y = 1-\frac{2.8}{x}$ that envelopes nearly 
all the sources unexpectedly falling below  S$_{\rm total}$/S$_{\rm peak}$ = 1.0 line. The reflection of this 
curve around  S$_{\rm total}$/S$_{\rm peak}$ = 1.0 line defined as $y = 1+\frac{2.8}{x}$ is plotted to account for the errors introduced by 
positive noise peaks. 
All the sources falling within the upper and lower curves $y = 1{\pm}\frac{2.8}{x}$ can be assumed as unresolved point sources,  
while extended sources lie outside the upper envelope curve. 
\par
From the total-to-peak flux density ratio versus SNR diagnostic plot, we find that all but seven of our radio sources 
are unresolved point sources (see Figure~\ref{size} ({\it Left panel})). 
We visually inspected all seven sources and find that only five sources appear marginally resolved in 
the 400~MHz uGMRT or 1.5 GHz JVLA images. 
We fitted resolved sources with two dimensional Gaussian using the JMFIT task in the Astronomical Image Processing 
System (AIPS\footnote{http://www.aips.nrao.edu/index.shtml}) that provides deconvolved sizes along with other 
parameters such as peak flux densities, integrated flux densities, and convolved sizes. 
We measure the radio sizes by considering the geometric mean of the deconvolved major axis and minor axis. 
Notably, deconvolved sizes of marginally resolved sources are of only a few arcsec. 
Thus, we find that the radio-detected DOGs in our sample exhibit mostly (104/109 $\sim$ 95.4$\%$) 
unresolved or marginally resolved radio emission in the 400~MHz uGMRT and 1.5 GHz JVLA observations.   
Hence, we conclude that, in general, radio emission in DOGs is likely to be compact such that 
it appears unresolved in the 400~MHz uGMRT and 1.5 GHz JVLA observations having resolution of 6$^{\prime\prime}$.0 and 4$^{\prime\prime}$.5, respectively. 
Our results are consistent with the findings of \cite{Gabanyi21} who reported that the IR-bright DOGs in their sample mostly exhibit 
unresolved radio emission in the 1.4 GHz FIRST observations with 5$^{\prime\prime}$.0 resolution.  
Further, five of our sample sources classified as resolved are only marginally extended and do not reveal radio structures. 
Thus, we find that the morphological details of our sample sources remained unclear due to the fact that only unresolved or 
slightly resolved emission is seen in our 400~MHz uGMRT and 1.5 GHz JVLA images.  
\par
We also attempted to constrain the physical sizes of the radio emission in our sample sources. 
Among 109 radio-detected sources only 50 sources (47 unresolved and three resolved sources) have available redshift 
estimates. For unresolved sources, we place an upper limit of 4$^{\prime\prime}$.5 on their angular sizes using 
the JVLA observations, whenever available, otherwise 6$^{\prime\prime}$.0 upper limit is used from the uGMRT observations. 
We use deconvolved radio sizes for slightly resolved sources.
Figure~\ref{size} ({\it Right panel}) shows the distribution of physical radio sizes or the upper limits placed on them for our sample sources.  
We find that the upper limits on physical sizes for 47/50 sources are distributed in the range of 29.4 kpc to 50.7 kpc 
with a median value of 37.5 kpc. The three marginally resolved sources have radio sizes of 2.7 kpc, 17.8 kpc and 26.8 kpc. 
From Figure~\ref{size} ({\it Right panel}) it is evident that the physical radio sizes in our sample DOGs are mostly smaller than 40 kpc 
and the radio$-$emitting structures are expected to reside within the host galaxy.
We note that recently \cite{Patil20} studied IR-bright radio-bright DOGs using 10 GHz JVLA observations of sub-arcsec resolution  
and found that nearly 93/155 (60$\%$) of their sample sources exhibit compact sizes of $<$0$^{\prime\prime}$.2 that corresponds to 1.7 kpc at 
$z$ = 2.0. Thus, we infer that, similar to radio-bright DOGs, our radio-faint (S$_{\rm 1.4~GHz}$ $<$1.0 mJy) DOGs powered by 
the AGN (see Section~\ref{nature}) can possess radio jets of a few kpc.  
\subsection{Radio spectral characteristics}
\label{sec:spectra}
\begin{figure*}[ht]
\includegraphics[angle=0,width=8.0cm,trim={0.0cm 9.0cm 0.0cm 6.5cm},clip]{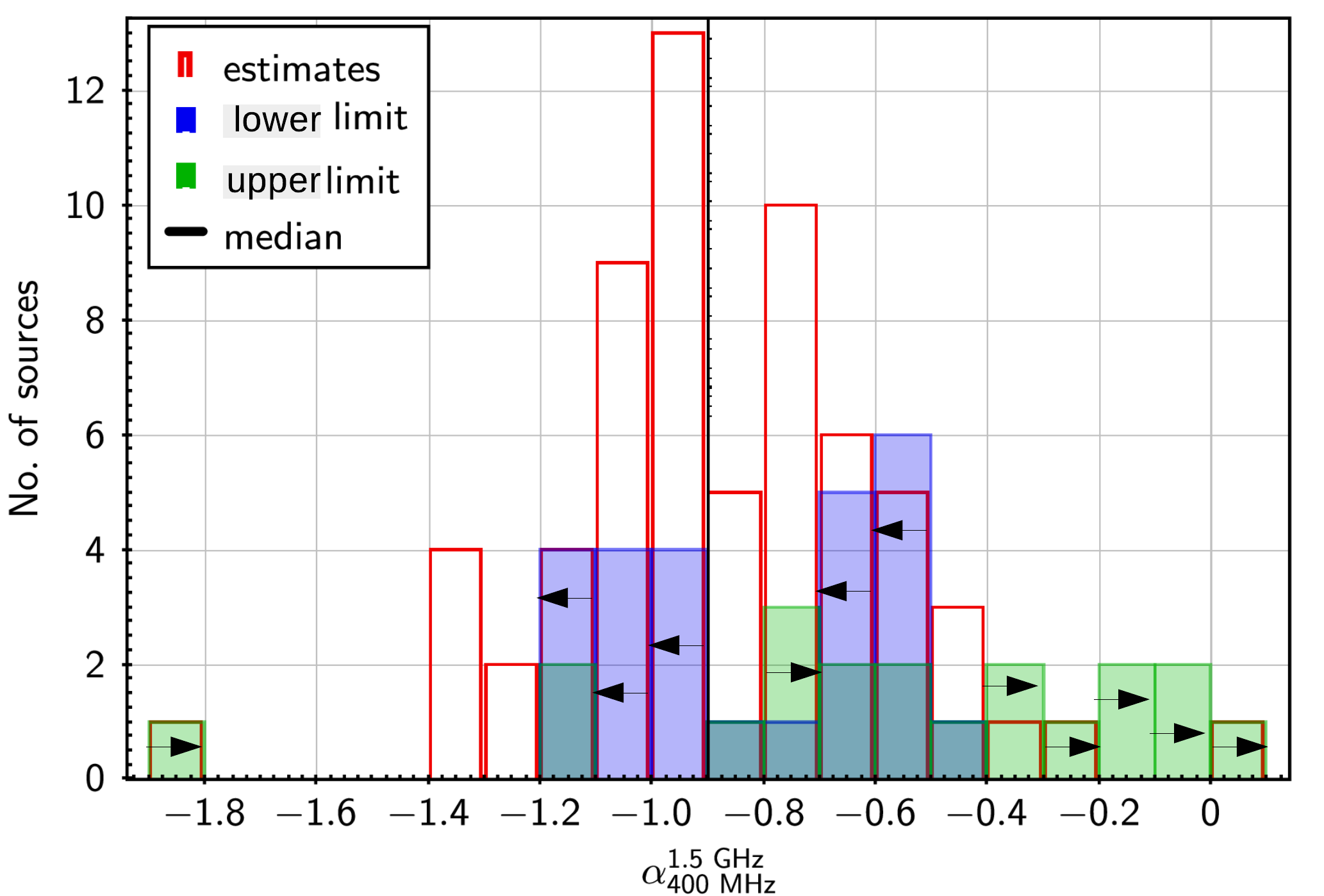}
{\includegraphics[angle=0,width=8.0cm,trim={0.0cm 6.5cm 0.0cm 6.5cm},clip]{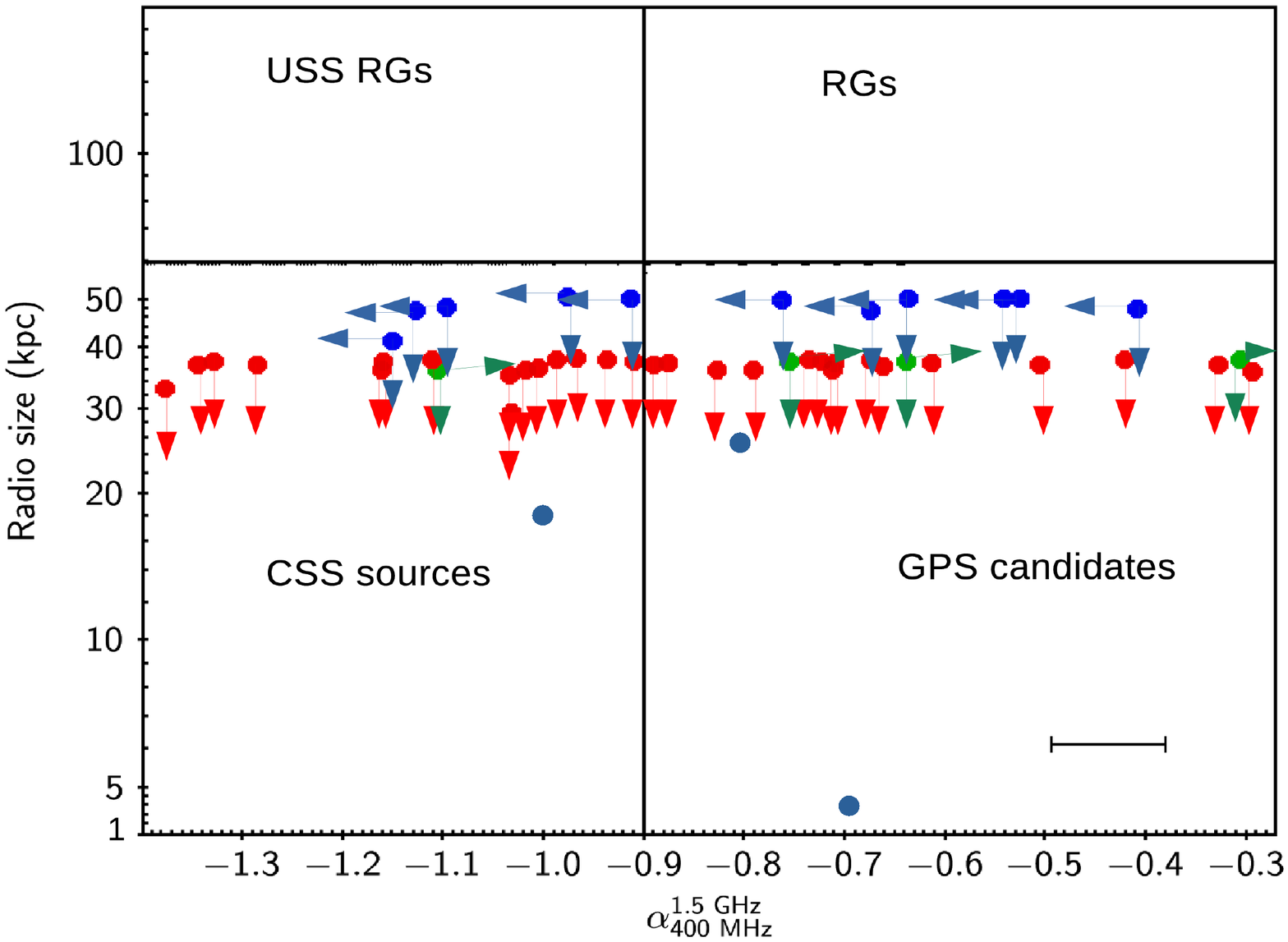}}
\caption{{\it Left panel} : The histogram of spectral index between 400 MHz and 1.5 GHz. {\it Right panel} : The radio sizes versus spectral indices plot. Sources with radio sizes and upper limits derived from the 400~MHz uGMRT and 1.5 GHz JVLA observations are indicated by Blue and Red points, respectively.}
\label{Alpha}
\end{figure*}

We examined the radio spectral indices of our sample sources using their total flux densities measured at 400 MHz and 1.5 GHz. Only 65/109 radio-detected sources have detections at both frequencies. 
We obtained an upper limit on the spectral indices of 26 sources detected at 400 MHz 
but not at 1.5 GHz. The 18 sources detected at 1.5 GHz but not at 400 MHz have only lower limits on their spectral indices. Figure~\ref{Alpha} ({\it Left panel}) shows the histogram of two-point spectral 
index measured between 400 MHz and 1.5 GHz (${\alpha}_{\rm 400~MHz}^{\rm 1.5~GHz}$). 
We find that the distribution of ${\alpha}_{\rm 400~MHz}^{\rm 1.5~GHz}$ spans in the range of -1.89 to +0.02 with 
a median value of -0.91. Notably, the median value is different than that ($\alpha$ = -0.7) usually seen 
for the general radio source population.
Figure~\ref{Alpha} ({\it Left panel}) shows that the spectral index (${\alpha}_{\rm 400~MHz}^{\rm 1.5~GHz}$) 
distribution for our sample sources is bimodal, {\ie}one set of sources show steep spectral index 
(${\alpha}_{\rm 400~MHz}^{\rm 1.5~GHz}$ $<$-0.9), while another set of sources shows a relatively 
flatter spectral index (${\alpha}_{\rm 400~MHz}^{\rm 1.5~GHz}$ $>$-0.8). The distributions of upper and lower limits 
on ${\alpha}_{\rm 400~MHz}^{\rm 1.5~GHz}$ also reinforce the bimodal trend. 
\par
\begin{figure}[ht]
\includegraphics[angle=0,width=8.0cm,trim={0.0cm 7.0cm 0.0cm 6.5cm},clip]{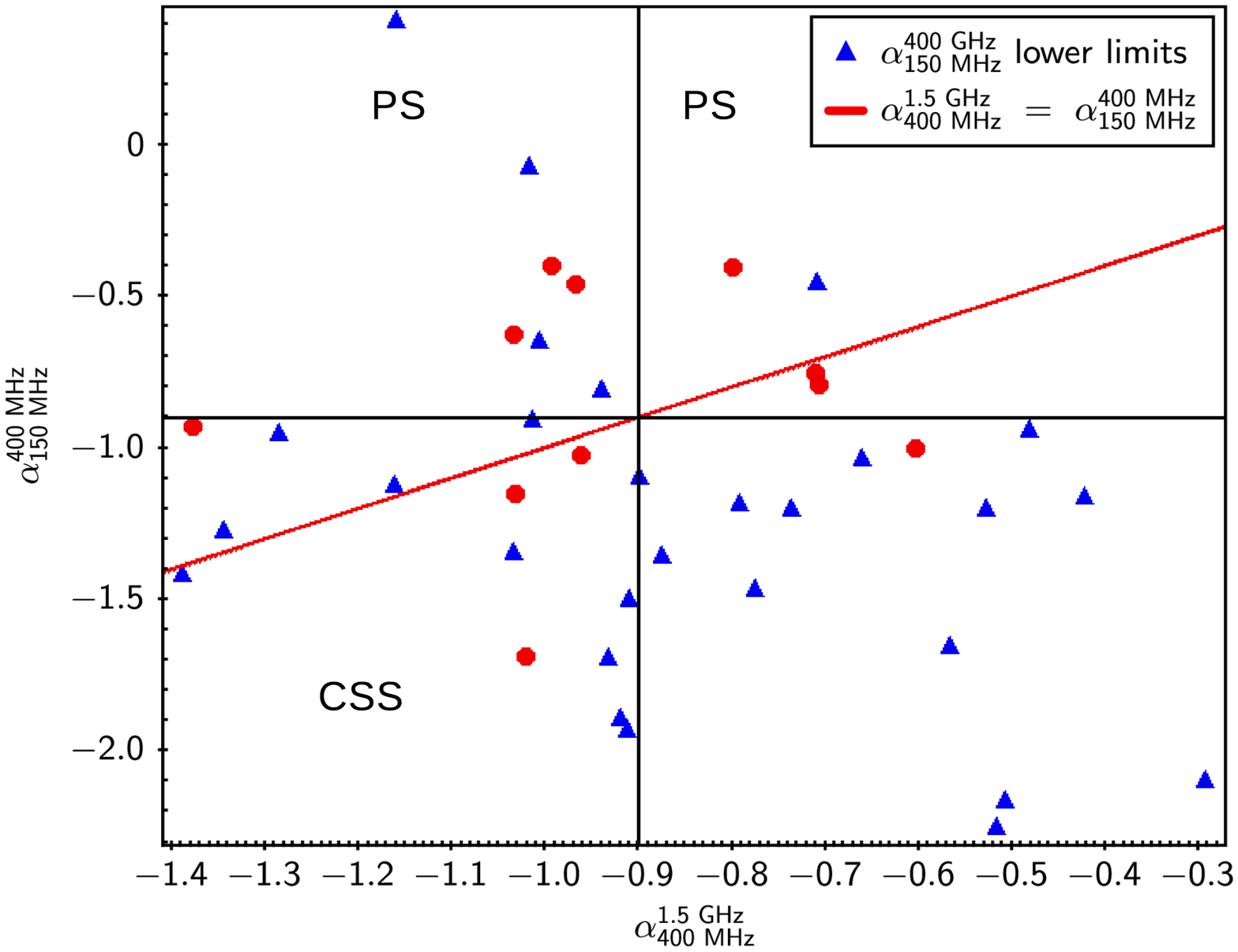}
\caption{The 400 MHz$-$1.5 GHz spectral index versus 150 MHz$-$400 MHz spectral index plot.}
\label{AlphaVsAlpha}
\end{figure}
To gain more insights into the nature of our sample 
sources we plot spectral indices versus physical radio sizes (see Figure~\ref{Alpha}, {\it Right panel}).     
It is clear that the steep spectrum sources are mostly compact, and hence, these can be classified as the 
Compact-Steep-Spectrum (CSS) sources. The CSS sources are known to exhibit the jet-lobe structure 
of a few kpc to a few tens of kpc with a steep spectrum ($\alpha$ $<$-1.0) 
and represent radio galaxies in their early phase of evolution \citep{ODea21}. 
Our remaining sample sources with non-steep spectral index are also compact (see Figure~\ref{Alpha}, {\it Right panel}). 
Given the small sizes but non-steep spectral indices, 
these sources can possibly constitute Peaked-Spectrum (PS) sources that are believed to represent the earliest phase 
of the evolution of radio galaxy with sizes $<$1.0 kpc. The PS sources exhibit a peak in their radio spectrum, 
and based on the peak frequency (${\nu}_{\rm p}$), they can be classified as the GHz-Peaked-Spectrum 
(GPS, 1.0~GHz$\leq$ ${\nu}_{\rm p}$ $\leq$5.0~GHz) sources, the High-Frequency-Peakers (HFP, ${\nu}_{\rm p}$ $>$5.0~GHz) 
and the MHz-Peaked-Spectrum (MPS, ${\nu}_{\rm p}$ $<$1.0 GHz) \citep[see][]{ODea21}. 
In Figure~\ref{Alpha} ({\it Right panel}), we classify compact non-steep spectrum sources as the PS candidates considering 
that the two-point spectrum is insufﬁcient to confirm the presence of peak in the radio spectrum. 
We note that the multi-frequency broadband radio SEDs are required to determine ${\nu}_{\rm p}$ to confirm their nature.       
\par
To examine the spectral behaviour at lower frequencies, we used 150 MHz flux densities, whenever available. 
Figure~\ref{AlphaVsAlpha} shows a plot of ${\alpha}_{\rm 400~MHz}^{\rm 1.5~GHz}$ versus  ${\alpha}_{\rm 150~MHz}^{\rm 400~MHz}$ 
in which only 65 sources with detection in at least two frequencies (400 MHz and 1.5 GHz) are considered. 
There are only 11/65 sources detected at all three frequencies (150 MHz, 400 MHz and 1.5 GHz) and have
estimates of ${\alpha}_{\rm 400~MHz}^{\rm 1.5~GHz}$ as well as ${\alpha}_{\rm 150~MHz}^{\rm 400~MHz}$. 
The majority (54/65) of our sources have only loose constraints on ${\alpha}_{\rm 150~MHz}^{\rm 400~MHz}$ lower limits 
due to a large mismatch between the sensitivities of uGMRT and LOFAR observations.   
From Figure~\ref{AlphaVsAlpha}, it is evident that the sources with no detection in the 150 MHz LOFAR observations 
can plausibly be PS sources that have 150 MHz flux densities much lower than the LOFAR detection limit.   
In fact, the sources classified as CSS on the basis of ${\alpha}_{\rm 400~MHz}^{\rm 1.5~GHz}$ can also be PS sources. 
A PS source at $z$ $\geq$ 1.2 (median redshift in our sample) with ${\nu}_{\rm p}$ = 400 MHz in the observed frame 
will have ${\nu}_{\rm p}$ $\geq$ 880 MHz in the rest frame. 
Thus, our study demonstrates that the DOGs in our sample are likely to host CSS and PS sources 
of small sizes, and we need deeper multi-frequency observations of high resolution (sub-arcsec or better) 
to ascertain this possibility. 
\subsection{Radio luminosities}
\label{sec:Lumin}
We examined the radio luminosity distributions of our sample sources.  
\begin{figure*}[ht]
\includegraphics[angle=0,width=8.0cm,trim={0.0cm 5.8cm 0.0cm 6.5cm},clip]{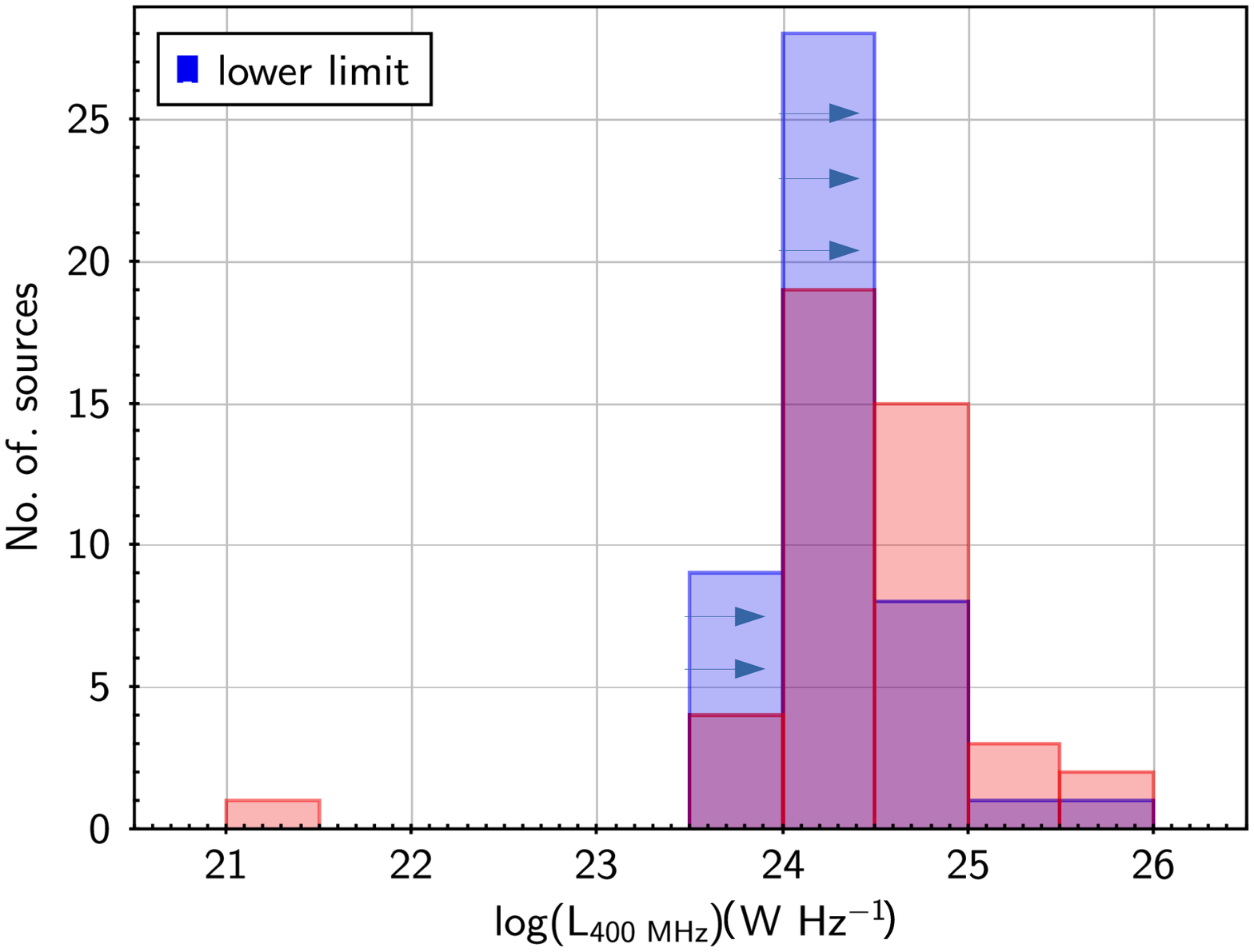}
{\includegraphics[angle=0,width=8.5cm,trim={0.0cm 6.8cm 0.0cm 7.0cm},clip]{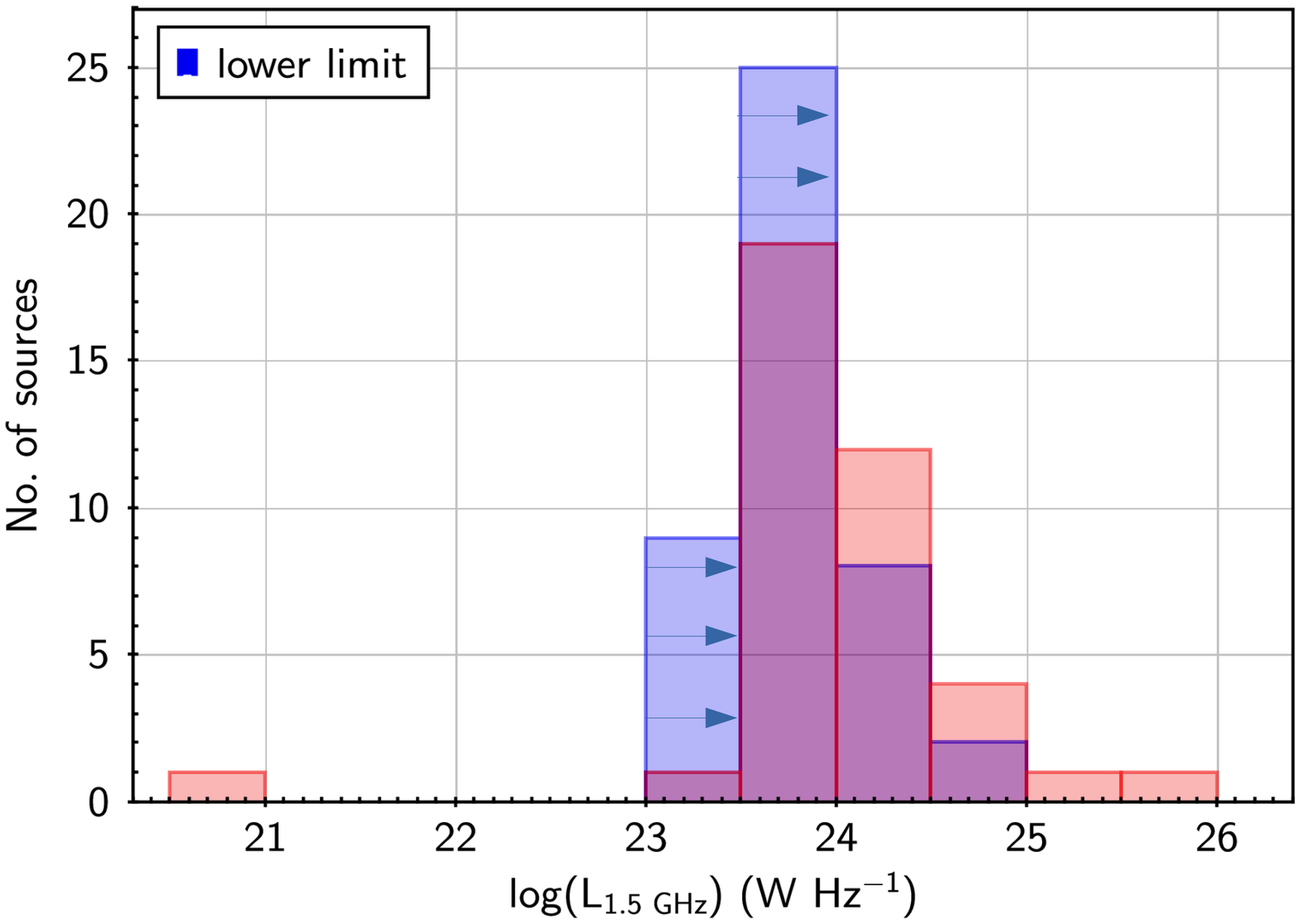}}
\caption{{\it Left panel} : The histogram of 400 MHz luminosity of our sample sources. {\it Right panel}: 
The histogram of 1.5 GHz luminosity of our sample sources. The lower limits on the radio luminosities are derived assuming 
the lower limit on their redshifts ($z$ $>$1.0).}
\label{Lumin}
\end{figure*}
%
%
%
We note that the radio luminosity estimates are limited only to 44/91 uGMRT detected sources and 39/83 JVLA detected sources 
that have available redshifts. For the remaining sources, we place a lower limit on their radio 
luminosities by assuming a 
conservative lower limit of $z$ $>$1.0 on their redshifts based on the empirical [3.6]~$-$~$z$ relationship, similar to the widely used $K-z$ relationship \citep{Willott03}.  
Figure~\ref{Lumin} shows the radio luminosity distributions at 400 MHz and 1.5 GHz. 
We find that the 400 MHz radio luminosities are distributed in the range of 2.3 $\times$ 10$^{21}$ W~Hz$^{-1}$ 
to 9.46 $\times$ 10$^{25}$ W~Hz$^{-1}$ with a median value of 2.81 $\times$ 10$^{24}$ W~Hz$^{-1}$. We find that the majority (39/44) 
of our sources have 400 MHz radio luminosity higher than 10$^{24}$ W~Hz$^{-1}$. 
The lower limits on 400 MHz radio luminosities are found to be distributed across 5.7 $\times$ 10$^{23}$ W~Hz$^{-1}$ 
to 3.7 $\times$ 10$^{25}$ W~Hz$^{-1}$ with a median value of 1.6 $\times$ 10$^{24}$ W~Hz$^{-1}$. 
Thus, the radio-detected sources with $z$~$>$1 are likely to be powerful radio sources. 
We note that the 1.5 GHz radio luminosity distribution spans over 9.1 $\times$ 10$^{20}$ W~Hz$^{-1}$ to 
3.3 $\times$ 10$^{25}$ W~Hz$^{-1}$ with a median value of 8.9 $\times$ 10$^{23}$ W~Hz$^{-1}$.  
The lower limits on 1.5 GHz luminosities are distributed over 2.4 $\times$ 10$^{23}$ W~Hz$^{-1}$ to 
7.5 $\times$ 10$^{24}$ W~Hz$^{-1}$ with a median value of 5.0 $\times$ 10$^{23}$ W~Hz$^{-1}$. 
We find that both 400 MHz and 1.5 GHz radio luminosity distributions exhibit a similar trend (see Figure~\ref{Lumin}).  
\par
We note that the high radio luminosity $>$10$^{24}$ W~Hz$^{-1}$ found in most of our sources 
can be considered as a signature of AGN originated radio emission. 
The relationship between radio luminosity and star formation rate 
L$_{\rm 1.4~GHz}$ (W~Hz$^{-1}$) = 4.6 $\times$ 10$^{21}$ [SFR (M$\odot$~yr$^{-1}$)], assuming the radio emission 
arising only from star-formation \citep{Condon02}, infers an extremely high star-formation rate 
of $>$100 M${\odot}$~yr$^{-1}$ required to produce the radio luminosity even for sources 
lying at the lower end of the luminosity distribution. 
Further, spectral characteristics, {\ie}CSS or potentially PS sources confirm the AGN originated radio emission. 
In Section~\ref{nature} we show that the MIR colour-colour diagnostic plot (see Figure~\ref{MIRWedge}) also confirms 
that our sample sources, including the radio-detected ones, are AGN dominated.  
We point out that the FIRST and NVSS selected radio-bright (S$_{\rm 1.4~GHz}$ $>$7 mJy) DOGs reported in \cite{Patil20} have 1.4 GHz radio luminosities in the range of 
10$^{25}$ W~Hz$^{-1}$ to 3.2 $\times$ 10$^{27}$ W~Hz$^{-1}$ with a median value of 2.0 $\times$ 10$^{26}$ W~Hz$^{-1}$. 
Thus, in comparison to the radio-bright DOGs, our radio-faint DOGs (S$_{\rm 1.5~GHz}$ $>$0.07 mJy) 
at the similar redshifts ($z_{\rm median}$ = 1.2) are nearly two dex less luminous, which can be understood due to their faintness in radio by a similar factor. 
Therefore, our deep 400~MHz uGMRT and 1.5 GHz JVLA observations unveil a population of less luminous 
radio-AGN in DOGs 
that were remained unexplored in the shallow radio surveys of the previous generation.  
\section{Prevalence of AGN in DOGs}
\label{nature}
Our radio observations have suggested that the radio-detected DOGs mostly contain AGN. 
To examine the prevalence of AGN in DOGs, we also study the MIR and X-ray properties of our sample sources.
\subsection{MIR colour-colour diagnostic}
\begin{figure}[ht]
\includegraphics[angle=0,width=8.0cm,trim={2.75cm 8.65cm 2.75cm 8.0cm},clip]{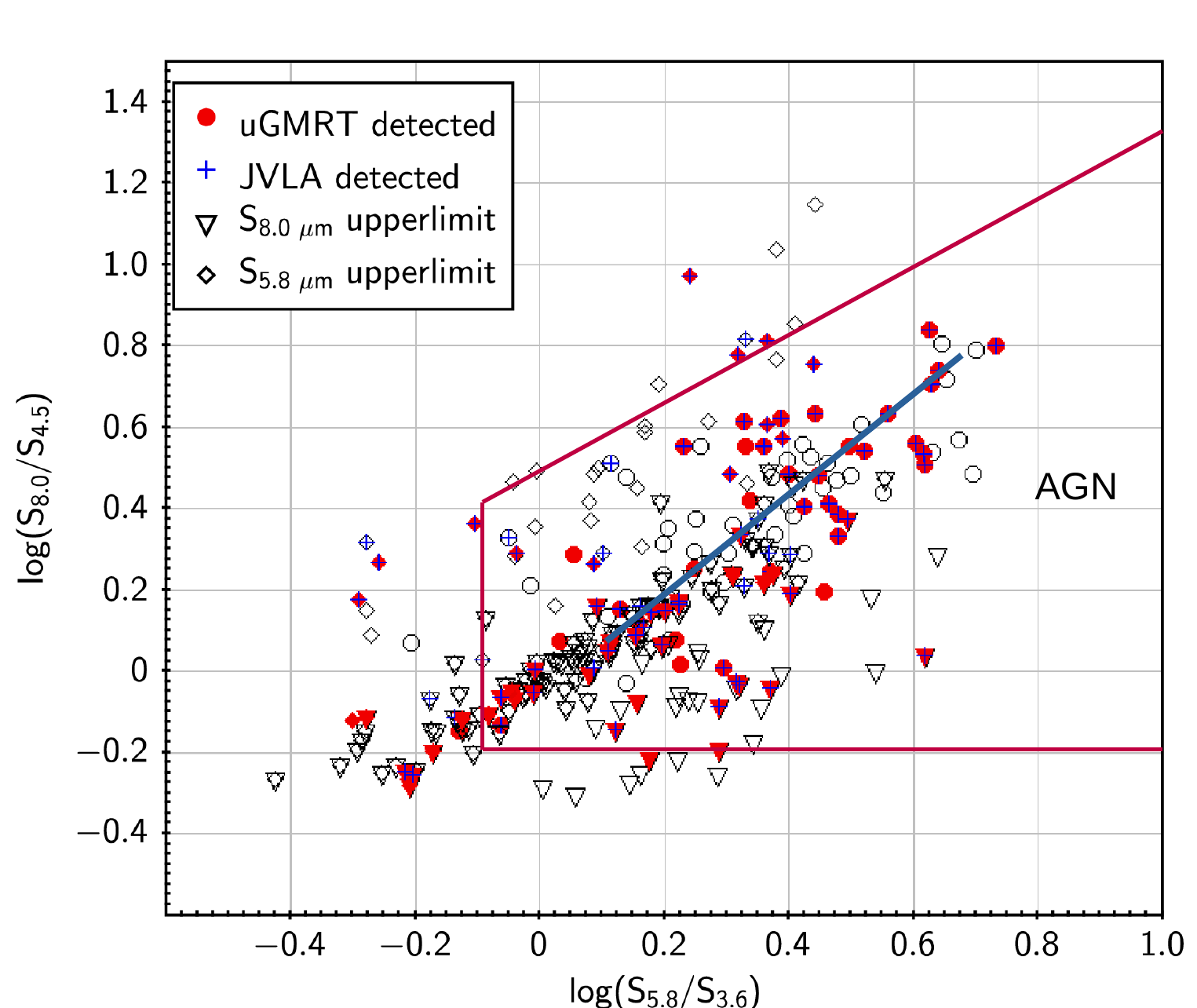}
\caption{The MIR colour-colour diagnostic plot. The AGN selection wedge proposed by 
\cite{Lacy04} is shown with the red solid lines depicting the boundaries of region defined as 
(log(S$_{5.8}$/S$_{3.6}$) $>$ -0.1) $\wedge$ (log(S$_{8.0}$/S$_{4.5}$) $>$ -0.2) $\wedge$ (log(S$_{8.0}$/S$_{4.5}$) $\leq$ 
0.8log(S$_{5.8}$/S$_{3.6}$) + 0.5); where $\wedge$ is `AND' operator. The blue diagonal line represents the locus of AGN showing power$-$law MIR emission with spectral index ranging over -0.5 to -3.0.}
\label{MIRWedge}
\end{figure}

The MIR colours based on the {\em Spitzer} IRAC observations have been used to identify the AGN population by 
exploiting the power$-$law spectral characteristics of AGN-heated dust emission in the MIR wavelengths 
\citep{Lacy04,Stern05,Donley12}. 
In fact, MIR colour criteria have proven efficient in identifying obscured-AGN that 
can remain elusive in the optical and X-ray wavelengths \citep[see][]{Fiore09}. 
In Figure~\ref{MIRWedge}, we show MIR colour-colour plot 
log(S$_{5.8}$/S$_{3.6}$) versus log(S$_{8.0}$/S$_{4.5}$); where S$_{3.6}$, S$_{4.5}$, S$_{5.8}$, and S$_{8.0}$ are fluxes 
at 3.6~$\mu$m,  4.5~$\mu$m,  5.8~$\mu$m and 8.0~$\mu$m bands, respectively. 
We find that only 268/321 of our MIPS-selected DOGs have IRAC coverage and fluxes available 
in the first two IRAC bands. There are only 
75/268 sources with fluxes available in all four IRAC bands. For sources lacking the detections 
in 5.8~$\mu$m or 8.0~$\mu$m bands, we used the upper limits of 27.5 $\mu$Jy and 32.5~$\mu$Jy 
on their fluxes, respectively.   
\par
Using the MIR colour-colour plot, we find that all but two of our DOGs (73/75 $\sim$ 97$\%$) 
with detections in four IRAC bands fall within the AGN selection wedge proposed by \cite{Lacy04}.  
Also, most of our DOGs are lying on or close to the power$-$law locus suggesting their MIR emission is mainly arising from AGN.
Sources with only upper limits on 5.8~$\mu$m or 8.0~$\mu$m fluxes also mostly fall within the AGN selection wedge 
(see~Figure~\ref{MIRWedge}). These sources are likely to remain within the AGN selection wedge if their fluxes are 
a few times lower than the 5$\sigma$ detection limits.         
Hence, from the MIR colour-colour diagnostic plot, it is evident that the vast majority of our sample DOGs are likely to contain AGN. 
In our sub-sample of 75 DOGs having MIR colour estimates, we find the radio detections in 37 sources, and all but one 
radio-detected sources fall within the AGN wedge. 
Hence, we can conclude that, a vast majority of radio-detected DOGs contain AGN, 
which is evident from the MIR colour-colour diagnostics and their radio characteristics.
Further, the MIR colour-colour plot also reveals a population of AGN-dominated DOGs with no radio emission. 
These sources can be radio-weak AGN that fall below the detection limit of our uGMRT and 
JVLA observations (see Section~\ref{sec:stack}). 
\subsection{X-ray emission in DOGs}
X-ray observations can allow us to measure the amount of obscuration in AGN.
Although, heavily obscured AGN known as the Compton-thick AGN (CT-AGN) with absorbing column density 
(N$_{\rm H}$) $\geq$ 1.5 $\times$ 10$^{24}$~cm$^{-2}$ 
are often missed in the {\em XMM-N} observations with energy coverage limited to 10 keV \citep{Ricci15}. 
The CT-AGN at higher redshifts can still be detected in the {\em XMM-N} observations owing to 
the shift of rest-frame high-energy ($E$ $>$10 keV) spectrum into the 0.5$-$10 keV band.  
We examine the X-ray detection of our sample DOGs by using the X-ray point-source catalogue 
obtained from the 1.3 Ms {\em XMM}-Newton observations \citep{Chen18} that were meant to cover the {\em Spitzer} 
Extragalactic Representative Volume Survey (SERVS) region and are found to overlap with the region covered by 
our uGMRT observations. 
With 46 ks median exposure time per pointing this survey reached down to the 5$\sigma$ flux limits 
of 1.7 $\times$ 10$^{-15}$ erg~cm$^{-2}$~s$^{-1}$, 1.3 $\times$ 10$^{-14}$ erg~cm$^{-2}$~s$^{-1}$, and 
6.5 $\times$ 10$^{-15}$ erg~cm$^{-2}$~s$^{-1}$ in the soft band (0.5-2.0 keV), hard band (2-10 keV), and full band (0.5-10 keV), 
respectively. 
We find that all but four of our sample sources fall within the {\em XMM}-SERVS coverage and 
only 24/317 (7.6$\%$) sources are detected in the X-ray.  
The 0.5$-$10 keV X-ray fluxes for our 24 sources are found in the range of 3.57 $\times$ 10$^{-15}$ erg~s$^{-1}$~cm$^{-2}$ to 
3.68 $\times$ 10$^{-14}$ erg~s$^{-1}$~cm$^{-2}$ with a median value of 1.8  $\times$ 10$^{-14}$ erg~s$^{-1}$~cm$^{-2}$ 
(see Table~\ref{tab:range}). 
We note that the measurement of absorbing column density (N$_{\rm H}$) requires X-ray spectral modelling, which may not 
be feasible for our very faint sources. Also, the detailed X-ray spectral modelling of X-ray detected DOGs 
is beyond the scope of this paper. 
To place constraints on N$_{\rm H}$, we use the hardness ratio (HR) parameter defined as H-S/H+S, where H and S 
are fluxes in the hard band (2.0-10 keV) and soft band (0.5-2.0 keV), respectively.    
\par
\begin{figure}[ht]
\includegraphics[angle=0,width=8.0cm,trim={0.0cm 6.5cm 0.0cm 6.5cm},clip]{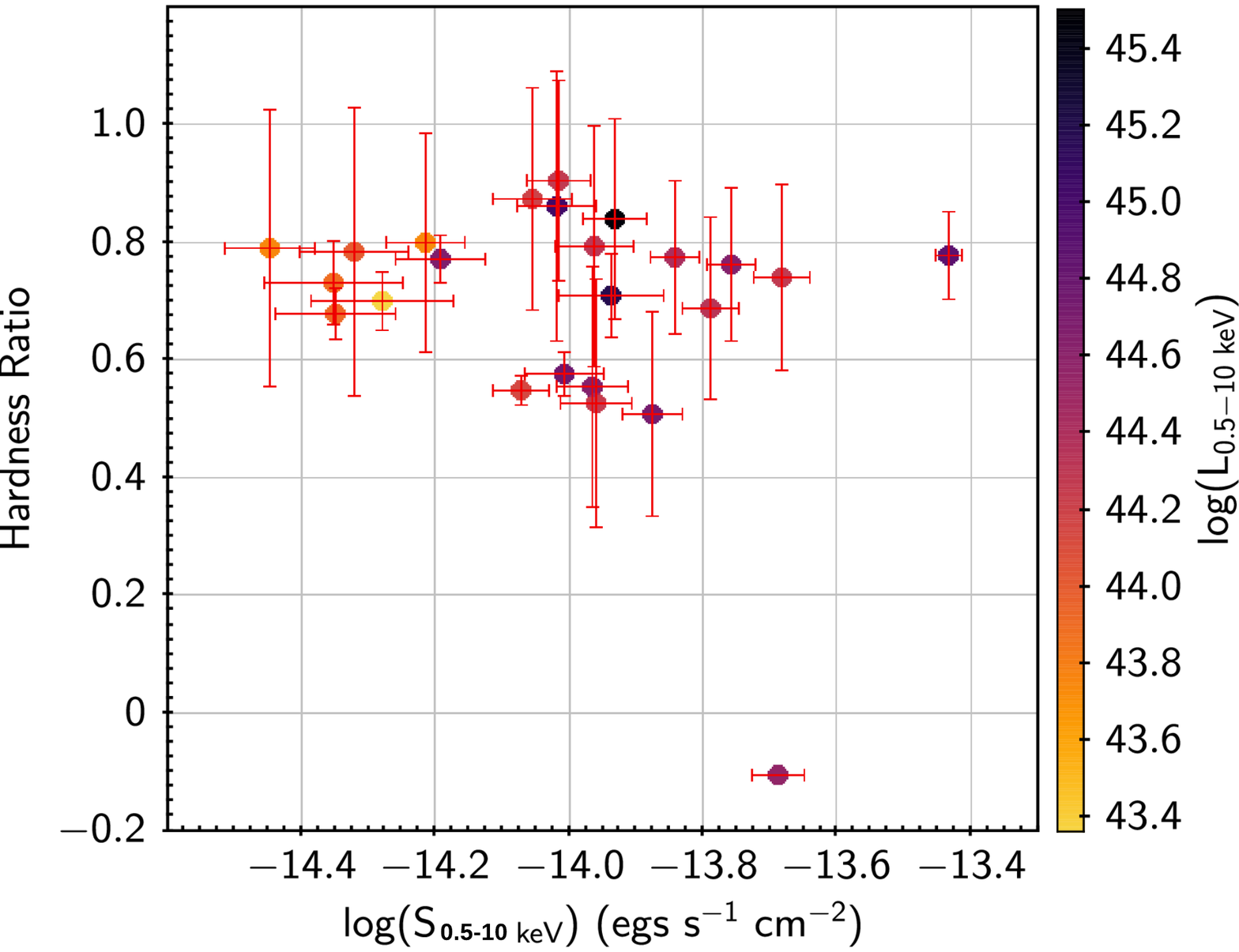}
\caption{The hardness ratio versus 0.5-10 keV X-ray flux.}
\label{X-ray}
\end{figure}

We use the N$_{\rm H}-$HR relationship plot \citep[from][]{Riguccini19} 
that predicts N$_{\rm H}$ from HR, assuming a power$-$law X-ray spectrum with photon index $\Gamma$ = 1.9 
and accounts for the effects of N$_{\rm H}$ and redshift ($z$) over a wide range of values. 
Figure~\ref{X-ray} shows the HR versus 0.5-10 keV flux plot for our sample sources. 
We find that the HR values of our sources are distributed in the range of -0.11 to 0.91, with a median value of 0.76.
In fact, all but one sources have HR $>$0.5. 
From the N$_{\rm H}-$HR relationship plot, we find that our sources 
are likely to have N$_{\rm H}$ higher than a few times of 10$^{23}$ cm$^{-2}$, hence, inferring them to be heavily obscured.
It is fairly possible that due to high obscuration and high redshift a significant population of AGN dominated DOGs may 
remain elusive in the existing X-ray observations. 
We note that all our X-ray detected sources have high 0.5$-$10 keV X-ray luminosity in the range of 
2.29 $\times$ 10$^{43}$ erg~s$^{-1}$ to 2.29 $\times$ 10$^{45}$ erg~s$^{-1}$ with a median value 
of 2.0 $\times$ 10$^{44}$ erg~s$^{-1}$. 
The observed X-ray luminosities of our sample sources are similar to those found for dust$-$obscured quasars 
\citep[see][]{Lansbury20}. 
Hence, the low X-ray detection rate 24/317 (7.6$\%$) in our sample DOGs even in the deep {\em XMM}-SERVS survey can be understood 
if less luminous as well as heavily obscured AGN are missed.
It is worth pointing out that the X-ray detection rate (7.6$\%$) in our sample DOGs is much lower than the radio 
detection rate (109/321 = 34$\%$). Therefore, our study demonstrates 
the importance of deep radio observations in unveiling the AGN population in DOGs that remained mostly 
undetected in the deep {\em XMM-N} X-ray survey.  
\section{Stacking Analysis}
\label{sec:stack}
Stacking is a useful tool to probe the existence of faint radio emission in sources 
with flux densities falling below the detection level of a given survey \citep{White07}. 
The noise-rms in stacked image decreases by a factor of $\sqrt N$, where 
$N$ is the number of image cutouts used for stacking, and all cutouts are assumed to be of the same depth.  
We used the stacking method to examine the existence of radio emission in the radio-undetected 
DOGs in our sample. 
We preferred median stacking over mean stacking as, unlike the mean, the median 
is less affected by extreme outliers. 
To obtain the deepest stacked image, we avoided the image cutouts of high 
noise-rms, {\ie}$>$40~$\mu$Jy~ beam$^{-1}$ and $>$20~$\mu$Jy~ beam$^{-1}$ 
in the 400~MHz uGMRT and 1.5 GHz JVLA 
images, respectively. We note that among 230/321 DOGs with no detected counterparts 
in the 400~MHz uGMRT observations, a large fraction 
of sources fall in the peripheral region having relatively high noise-rms ($>$40 $\mu$Jy~ beam$^{-1}$), 
and only 116 sources lying mostly in the central region have noise-rms $<$40~$\mu$Jy~ beam$^{-1}$. 
The 1.5 GHz JVLA mosaiced image of several pointings has nearly constant 
noise-rms, and 206/238 DOGs with no 1.5 GHz detected counterparts have noise-rms $<$20~$\mu$Jy~ beam$^{-1}$. 
We stacked 1$^{\prime}$ $\times$ 1$^{\prime}$ radio image cutouts centred at the optical positions of 
our radio-undetected DOGs and computed the pixel-by-pixel median value. 
\par 
\begin{figure*}[h]
\includegraphics[angle=0,width=8.0cm,trim={0.0cm 0.0cm 0.0cm 0.0cm},clip]{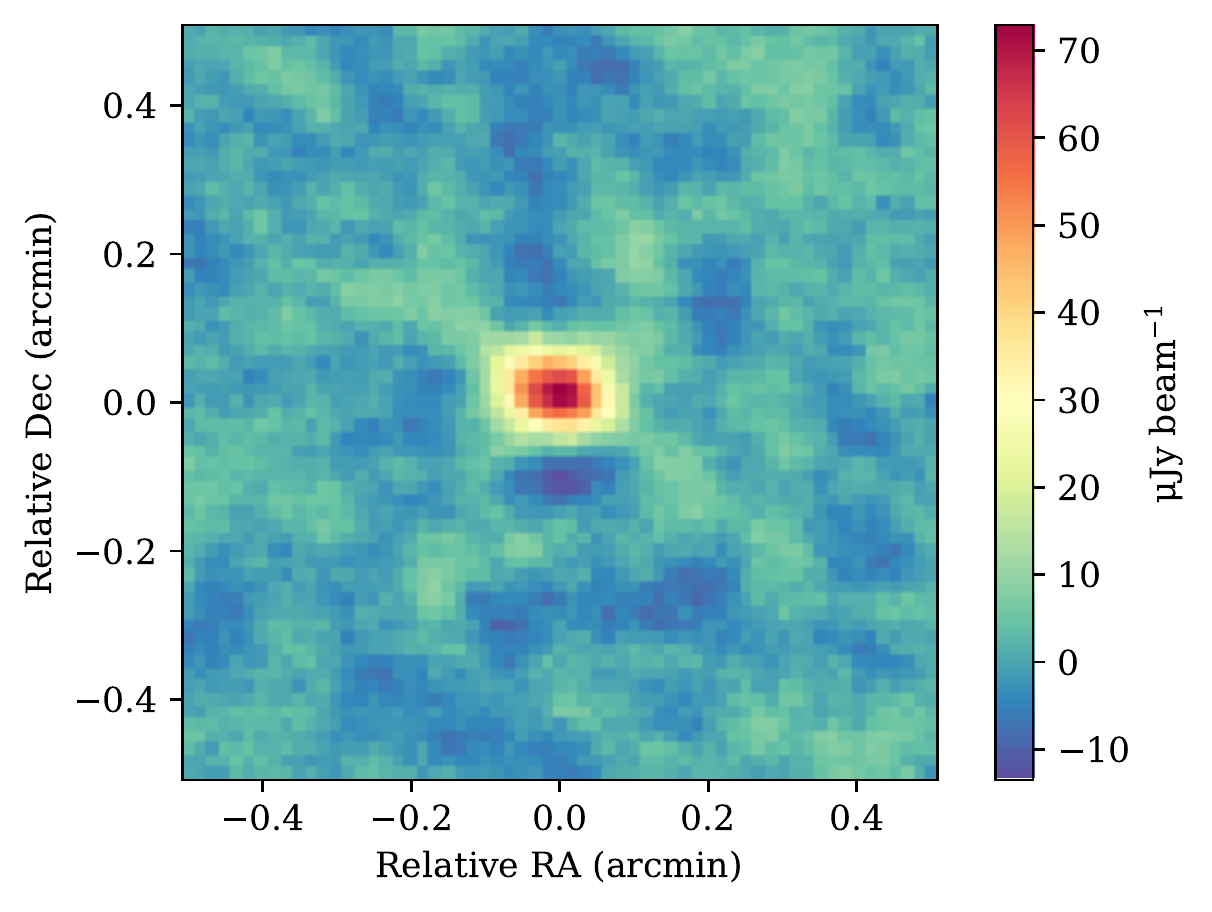}
{\includegraphics[angle=0,width=8.0cm,trim={0.0cm 0.0cm 0.0cm 0.0cm},clip]{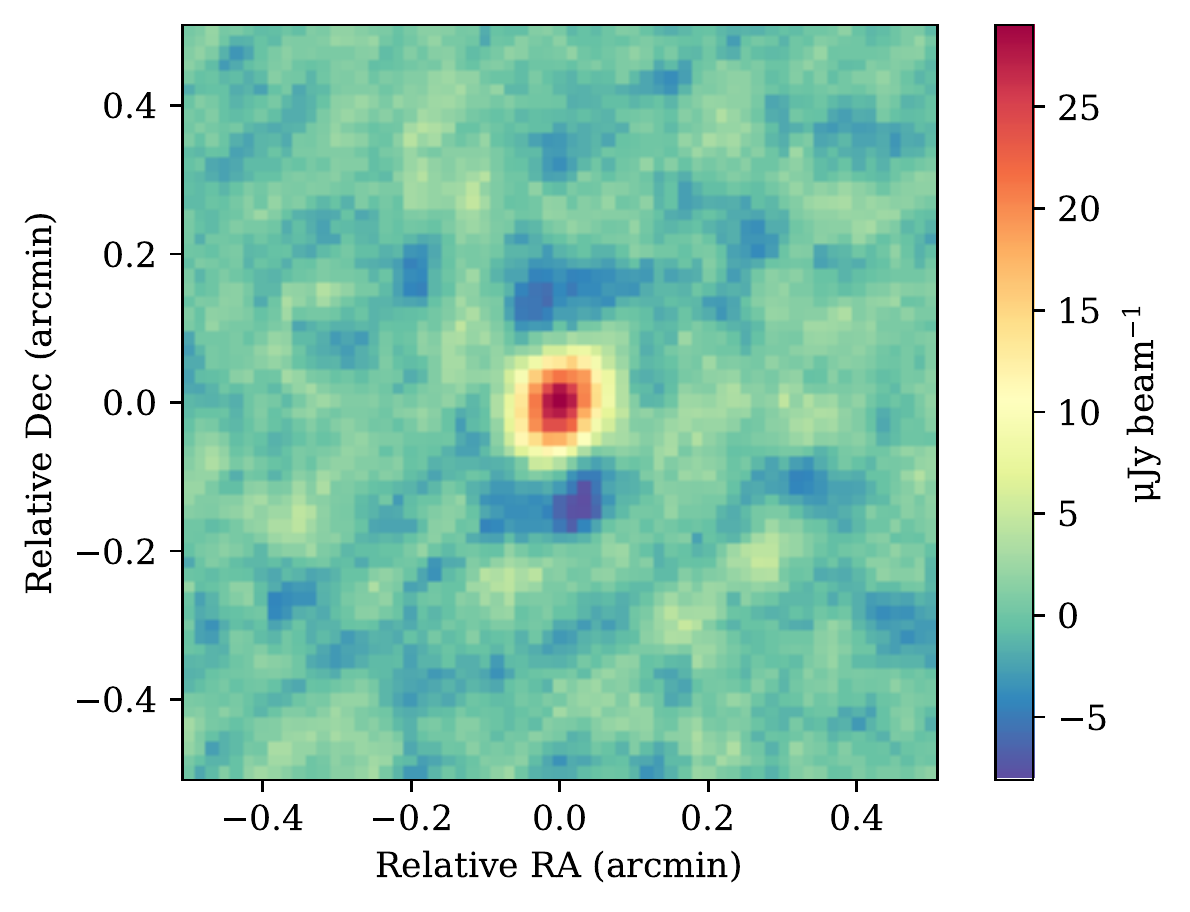}}
\includegraphics[angle=0,width=8.0cm,trim={0.0cm 0.0cm 0.0cm 0.0cm},clip]{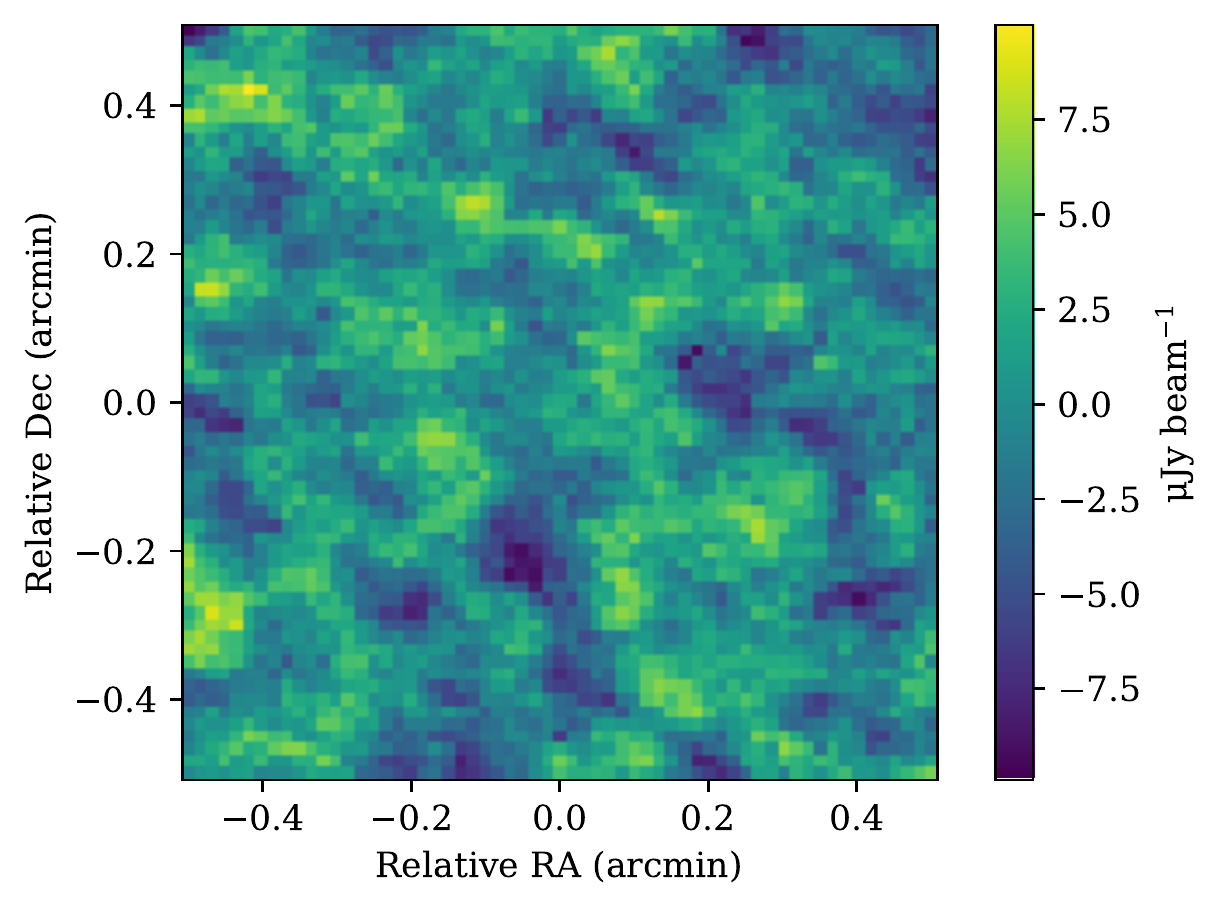}
\hspace{1.5cm}
{\includegraphics[angle=0,width=8.0cm,trim={0.0cm 0.0cm 0.0cm 0.0cm},clip]{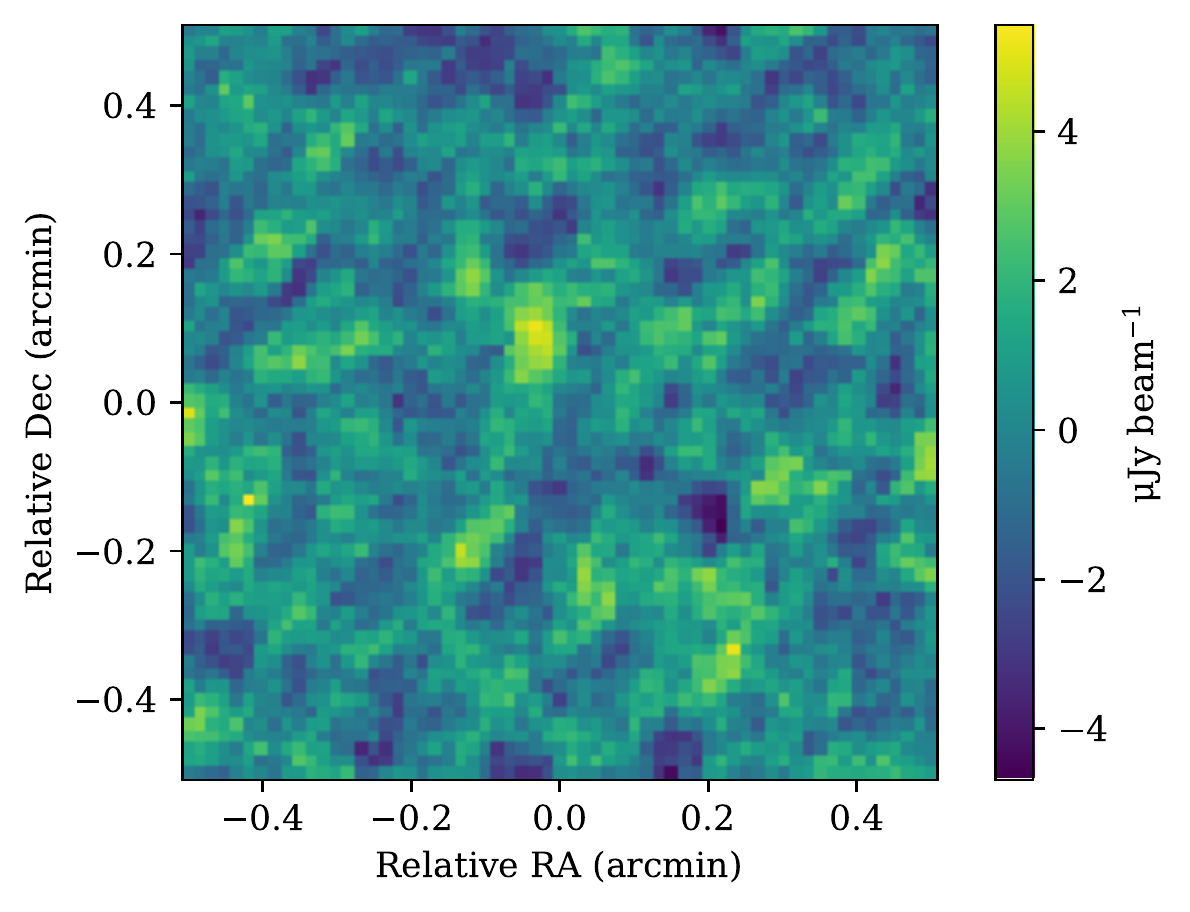}}
\caption{The median-stacked images of DOGs with no detected radio counterparts. 
{\it Upper left panel}: The 400~MHz uGMRT stacked image with 
cutouts centred at the DOGs positions. {\it Upper right panel}: The 1.5 GHz JVLA stacked image with 
cutouts centred at the DOGs positions. {\it Lower left panel}: The 400~MHz uGMRT stacked image with 
cutouts centred at random positions. {\it Lower right panel}: The 1.5 GHz JVLA stacked image with 
cutouts centred at random positions.}
\label{fig:stack}
\end{figure*}
\begin{table*}[h]
\caption{The details of stacked 400 MHz uGMRT and 1.5 GHz JVLA images.}
\label{tab:stack} 
\hspace{1.0cm}
\begin{tabular}{lcccccccc}
\topline
Observations & N$_{\rm cutouts}$ & S$_{\rm peak}$  &  noise-rms & SNR  & Size & PA &  Positions \\  
          &       & (${\mu}$Jy~beam$^{-1}$) & (${\mu}$Jy~beam$^{-1}$) &   & ($^{\prime\prime}$~$\times$~$^{\prime\prime}$) & (deg) &    \\ \midline
 400~MHz uGMRT & 116   &   72.9   &  3.4  & 21.4  & 7.7$\times$5.2 & 82   &  DOGs  \\
             & 116   &   10.8   &  3.1  & 3.5   &                &      &  random   \\
1.5 GHz JVLA & 206   &   29.0   &  1.5  & 19.3  & 6.2$\times$5.1 & 137  &  DOGs     \\ 
             & 206   &   5.2    &  1.4  & 3.7   &                &      &  random  \\ 
\hline
\end{tabular}
\tablenotes{{\it Notes} - The cutoff on noise-rms $<$40 $\mu$Jy~ beam$^{-1}$ and $<$20 $\mu$Jy~ beam$^{-1}$
is applied for the 400~MHz uGMRT and 1.5~GHz image cutouts, respectively. The off-centered strongest signal 
(brightest pixel) is considered in the median-stacked images for random positions.} 
\end{table*}
In Table~\ref{tab:stack}, we list our stacking results. We note that the 400~MHz uGMRT median-stacked 
image shows a clear detection with the brightest central pixel value 
corresponding to the flux density of 72.9 $\mu$Jy~beam$^{-1}$ and SNR 21.4 
(see Figure~\ref{fig:stack}, {\it upper left panel}).   
The noise-rms of 400~MHz median-stacked image is only 3.4~$\mu$Jy~beam$^{-1}$,  
which is, as expected, nearly $\sqrt{N}$ times lower than the median noise-rms 
(30~$\mu$Jy~beam$^{-1}$) of our uGMRT observations. 
Hence, stacking analysis allows us to probe radio emission at a level which is nearly 
ten times fainter than that is directly detected in our uGMRT observations.  
Further, to ensure the reliability of detection in our stacked image, we performed stacking of radio 
image cutouts centred at the positions randomly shifted 
by 1$^{\prime\prime}$$-$2$^{\prime\prime}$ {\wrt}DOGs positions. 
We find that the median-stacked image with random positions yields no detection of any emission 
significantly higher than the noise-rms of 3.1~$\mu$Jy~beam$^{-1}$ 
(see Figure~\ref{fig:stack}, {\it lower left panel}). 
In fact, the brightest pixel lying far away from the centre in 1$^{\prime}$~$\times$~1$^{\prime}$ stacked image, 
corresponds to the flux density of 10.8~$\mu$Jy~beam$^{-1}$ with SNR 3.5. Hence, we find that the 
radio emission detected in our uGMRT median-stacked image, centred at the positions of radio-undetected DOGs, 
is genuine. 
\par
A similar exercise is performed to obtain the median-stacked image at 
1.5 GHz with the JVLA image cutouts (see Figure~\ref{fig:stack}, {\it upper right panel}). 
We find a clear detection with flux density 29.0~$\mu$Jy~beam$^{-1}$ and SNR 19.3 above the 
noise-rms of 1.5~$\mu$Jy~beam$^{-1}$ (see Table~\ref{tab:stack}). 
The 1.5~GHz stacked image at random positions shows only noise with no significant emission 
(see Figure~\ref{fig:stack}, {\it lower right panel}). 
We note that, similar to 400~MHz uGMRT stacked image, the 1.5 GHz stacked image 
(noise-rms $\sim$ 1.5~$\mu$Jy~beam$^{-1}$) is also nearly ten times deeper than 
the JVLA image (noise-rms $\sim$ 16~$\mu$Jy~beam$^{-1}$).   
Thus, using stacking analysis we demonstrate that our sample DOGs with no directly detected 
counterparts in the 400~MHz uGMRT and 1.5 GHz JVLA observations, possess faint radio emission with peak 
flux density of 72.9~$\mu$Jy~beam$^{-1}$, 29~$\mu$Jy~beam$^{-1}$ at 400 MHz and 1.5 GHz, respectively.
It is worth noting that both 400~MHz uGMRT and 1.5 GHz JVLA median-stacked images show 
nearly an unresolved point source emission (see Figure~\ref{fig:stack}, {\it upper panel}). 
The central emission components detected in the stacked images are fitted with a single Gaussian 
and gives sizes of 7$^{\prime\prime}$.7~$\times$~5$^{\prime\prime}$.2 and 
6$^{\prime\prime}$.2~$\times$~5$^{\prime\prime}$.1, 
that are similar to the synthesized beam sizes of  uGMRT 
(6$^{\prime\prime}$.7~$\times$~5$^{\prime\prime}$.3) and 1.5 GHz JVLA 
(4$^{\prime\prime}$.5~$\times$~4$^{\prime\prime}$.5) observations. 
Thus, we find that, similar to our radio-detected DOGs, radio-undetected DOGs are likely to 
possess compact radio emission smaller than 6$^{\prime\prime}$.0 at much fainter levels. 
\par
Further, to understand the nature of DOGs with feeble radio emission but 
no directly detected radio counterparts in our 400~MHz uGMRT and 1.5~GHz JVLA observations, 
we compare their redshift distributions and average 
radio luminosities with the radio-detected DOGs. 
We note that only 50/109 (45.9$\%$) radio-detected DOGs have estimated redshifts distributed 
across 0.034 to 2.64 with a median value of 1.22, while, only 81/212 (38.2$\%$) radio-undetected 
DOGs have redshifts in 
the range of 0.693 to 2.99 with a median value of 1.19. Although, the median redshifts of both 
sub-populations appear similar, but a lower fraction of sources with available redshifts 
indicates that the radio-undetected DOGs possibly reside at higher redshifts than the 
radio-detected ones. 
Further, for radio-undetected sources, we find an average 
L$_{\rm 1.5~GHz}$ = 2.25 $\times$ 10$^{23}$ W~Hz$^{-1}$ using S$_{\rm 1.5~GHz}$ = 0.029 
mJy~beam$^{-1}$ derived from the median-stacked image and median redshift of 1.19. 
We note that the average 1.5 GHz luminosity can only be a lower limit considering that a large 
fraction of potentially high$-z$ sources lack redshift estimates. 
Moreover, the average radio luminosity of otherwise radio-undetected DOGs infers 
them to be powered by a 
low-luminosity AGN or high star-formation rate ($\sim$100 M$\odot$~yr$^{-1}$) or both. 
We emphasize that more sensitive radio observations are required to unveil their nature 
via direct detections of such individual sources. 
In the next section, we discuss the importance of planned and upcoming deeper radio continuum surveys.
\section{Importance of the deeper radio continuum surveys from the SKA and its pathfinders}
\label{SKA}
We note that deep radio continuum surveys are planned with the upcoming largest radio telescope array SKA operating 
at various frequencies, {\ie}50 MHz $-$ 350 MHz, and 350 MHz $-$ 15.3 GHz frequency domains 
covered with the SKA1-low and the SKA1-mid, respectively. 
As per the baseline design performance, SKA1-mid is expected to achieve noise-rms of 
4.4 $\mu$Jy beam$^{-1}$ in one hour observing time with an angular resolution of 0$^{\prime\prime}$.7 in the 
0.35$-$1.05 GHz band centred at 770 MHz 
(see SKA factsheet\footnote{https://www.skatelescope.org/wp-content/uploads/2018/08/16231-factsheet-telescopes-v71.pdf}).  
We also note that, prior to the SKA surveys, deep radio continuum surveys although of smaller sky-area coverage, have already been conducted with the SKA pathfinder telescopes, such as the MeerKAT in South Africa and 
the Australian SKA pathfinders (ASKAP). 
The MeerKAT International GHz Tiered Extragalactic Exploration (MIGHTEE) survey 
achieved 2.0~$\mu$Jy beam$^{-1}$ noise-rms at 950~MHz $-$1.7 ~GHz band (central frequency 1284 MHz) 
with a total sky coverage of 20~deg$^{2}$ in various deep fields \citep[see][]{Heywood22}. 
The Evolutionary Map of the Universe (EMU) survey conducted with the ASKAP at 
944 MHz offers continuum images with noise-rms of 25$-$30~$\mu$Jy~beam$^{-1}$ and resolution 
of 11$^{\prime\prime}$ $-$ 18$^{\prime\prime}$ over a large area of 270 deg$^{2}$ \citep{Norris21}.
\par
We emphasize that, in comparison to our 400~MHz uGMRT (noise-rms = 30~$\mu$Jy beam$^{-1}$) 
and 1.5 GHz JVLA (noise-rms = 16~$\mu$Jy beam$^{-1}$) observations, the MIGHTEE survey provides 
deeper radio images. Also, the EMU survey with its large sky-area coverage (270 deg$^{2}$) and 
depth similar to the 1.5 GHz JVLA observations, would allow us to explore a much larger 
source population. As expected, a much deeper SKA1-mid survey with sub-arcsec angular resolution 
would be more useful in revealing compact radio structures in DOGs. 
For instance, most of the DOGs in our sample appeared unresolved in our 400~MHz uGMRT and 
1.5 GHz JVLA observations with the angular resolution of 4$^{\prime\prime}$.5$-$6$^{\prime\prime}$.0, 
but, deep SKA-mid observations of sub-arcsec resolution can show jet-lobe structure piercing 
through the dense environments of their host galaxies at $z$ $=$ 1$-$2. 
Also, SKA observations spanning across a wide range of frequencies would enable us to 
measure ${\nu}_{\rm p}$ and confirm the evolutionary stage of radio sources hosted in DOGs 
which are inferred to host CSS or PS sources.  
With stacking, we confirmed the existence of faint radio emission 
(72.9~$\mu$Jy~beam$^{-1}$ at 400 MHz and 29~$\mu$Jy~beam$^{-1}$ at 1.5~GHz) in otherwise 
radio-undetected sources in our 400~MHz uGMRT and 1.5 GHz JVLA observations. 
We note that the deeper radio surveys from the MIGHTEE and SKA would directly detect faint 
radio emission in DOGs that remained undetected in the current radio observations. 
Therefore, our study presents a demonstrated science case for the deeper radio continuum surveys 
from the SKA and its pathfinders.   
\par
Further, we emphasize that Very Long Baseline Interferometric (VLBI) observations providing 
milli$-$arcsecond (mas) resolution can enable us to detect radio structures at parsec$-$scales. 
It is worth {mentioning} that \cite{Frey16} performed VLBI observations using the European VLBI Network 
(EVN\footnote{https://www.evlbi.org/}) at 
1.7 GHz for four Hot DOGs at $z$ $\sim$ 1.7 $-$ 2.6. All four sources appearing unresolved in the FIRST observations 
have a total 1.4~GHz flux density in the range of 1.4 mJy $-$ 3.6 mJy. The EVN observations 
revealed extended radio emission with their projected linear sizes of hundreds to thousands of parsec. 
Notably, in two of their sample sources, \cite{Frey16} found that the flux density of the VLBI$-$detected component 
is much lower than the total flux density observed in the FIRST, suggesting that 70$-$90 per cent of 
the radio emission originates from angular scales larger than that probed by the EVN. 
Recently, \cite{Fan20} performed VLBI observations of a hyper-luminous dust-obscured quasar W2246$-$0526 
at $z$ = 4.6 using EVN plus enhanced Multi$-$Element Radio Linked Interferometer 
Network (e-MERLIN\footnote{https://www.e-merlin.ac.uk/index.html}) and the Very Long Baseline Array 
(VLBA\footnote{https://science.nrao.edu/facilities/vlba}), and detected only an unresolved 
nuclear component ($<$32 pc) at 1.66 GHz in the EVN plus e-MERLIN observations, while this source 
remained undetected in the VLBA observations. Also, flux density of the core component 
(75$\pm$9 ${\mu}$Jy) detected in the EVN plus e-MERLIN observations accounts for only about 10 per cent of 
that detected in the FIRST. Therefore, considering the fact that VLBI detection requires compact mas-scale core 
emission of nearly 0.1 mJy, and up to 90 per cent flux density may arise from the region larger than mas-scales, we can expect that only relatively bright DOGs ({\eg}S$_{\rm 1.4~GHz}$ $\geq$ 1.0 mJy) can possibly be regarded as good candidates for the VLBI observations.  
Hence, we emphasize that the relatively radio-bright DOGs reported in our study 
may also be used for VLBI observations to probe the early evolutionary stage of radio jets 
residing in dusty environments.    
\section{Summary and conclusions}
\label{Conclusion}
Using sensitive 400~MHz uGMRT (5$\sigma$ = 0.15 mJy beam$^{-1}$) and 1.5 GHz JVLA 
(5$\sigma$ = 0.08 mJy beam$^{-1}$) 
observations, we investigated the radio emission characteristics of 321 DOGs selected with the main criterion 
of flux ratio of 24~$\mu$m to $r$ band optical (S$_{\rm 24~{\mu}m}$/S$_{r}$) $\geq$1000. 
The combination of deep 24~$\mu$m {\em Spitzer} and HSC-SSP optical 
data allows us to select IR-faint (S$_{\rm 24~{\mu}m}$ $\geq$ 0.4 mJy) DOGs at higher redshift (z$_{\rm median}$ $\geq$1.2). 
The main conclusions of our study are outlined below.
\begin{itemize}
\item Our 400~MHz uGMRT observations detect only 91/321 (28.4 per cent) of our sample sources, 
and yield the highest detection 
rate, in comparison to the 1.5 GHz JVLA and 150 MHz LOFAR observations. With the combination of 400~MHz 
uGMRT and 1.5 GHz JVLA observations, the radio detection rate increases to 109/321 (34 per cent), if detection at one frequency 
(400~MHz uGMRT or 1.5 GHz JVLA) is considered. We note that the radio detection rates in the uGMRT and JVLA observations are 
much higher than that ($<$1.0 per cent) obtained with the FIRST and VLASS. 
The higher detection rates in the uGMRT and JVLA can be attributed to their better sensitivities 
and the radio emission of sub-mJy level in our sample DOGs. 
\item We find that our sample DOGs mostly have radio emission at sub-mJy level with 400 MHz flux densities 
distributed in the range of 0.137 mJy to 26.9 mJy with a median value of 0.37 mJy. The flux densities at 1.5~GHz are found in the range of 0.035 mJy to 9.33 mJy with a median value of 0.13 mJy. Notably, 79/83 (95 per cent) JVLA 
detected sources have 1.5 GHz flux density less than 1.0 mJy.    
\item In our uGMRT and JVLA observations, all except five sources appear unresolved with angular sizes $<$4$^{\prime\prime}$.5. 
Five sources are only marginally resolved and do not reveal any morphological details. 
The distribution of upper limits on the radio sizes suggests that most of our sources are 
expected to possess compact radio emission ($<$40 kpc at $z$ = 1.2) confined within their hosts.      
\item The distribution of two-point spectral index (${\alpha}_{\rm 400~MHz}^{\rm 1.5~GHz}$) appears bimodal 
with most sources having either steep (${\alpha}_{\rm 400~MHz}^{\rm 1.5~GHz}$ $<$-0.9) or 
flat (${\alpha}_{\rm 400~MHz}^{\rm 1.5~GHz}$ $>$-0.8) spectral index. The spectral index and upper limits derived 
at the lower-frequencies using 150 MHz LOFAR observations suggest that our sources are potentially CSS or PS sources.  
Thus, considering their compact radio sizes and spectral characteristics, we infer that most of our radio sources in DOGs 
are likely to be CSS or PS representing the young radio sources residing in obscured environments. 
Our results are consistent with the findings of \cite{Patil20}, who reported the existence of 
young compact radio-jets in radio-bright DOGs. 
\item We find that both 400 MHz and 1.5 GHz radio luminosity distributions span in the range of 10$^{21}$ W~Hz$^{-1}$ to 
10$^{26}$ W~Hz$^{-1}$ with a median value of nearly 10$^{24}$ W~Hz$^{-1}$. In fact, all but one of our sample sources have 
1.5~GHz luminosities higher than 10$^{23}$ W~Hz$^{-1}$, suggesting them to be mainly powered by AGN.
\item The MIR colour-colour diagnostic plot also confirms that the most of our radio-detected sources contain AGN. 
Further, many radio-undetected sources too fall within the MIR AGN selection wedge, revealing a population of AGN that remained undetected in our deep 400~MHz uGMRT and 1.5 GHz JVLA observations. 

\item The deep {\em XMM}-SERVS survey could detect only 24/317 (7.9 per cent) of our sample sources. 
The absorbing column densities (N$_{\rm H}$) estimated from the hardness ratios suggest that all our X-ray detected sources are 
heavily obscured with N$_{\rm H}$ $>$10$^{23}$ cm$^{-2}$. The observed 0.5$-$10 keV X-ray luminosities distributed 
in the range of 2.29 $\times$ 10$^{43}$ erg~s$^{-1}$ to 2.29 $\times$ 10$^{45}$ erg~s$^{-1}$ with a median value 
of 2.0 $\times$ 10$^{44}$ erg~s$^{-1}$, are similar to the luminous quasars. We infer that the {\em XMM}-SERVS survey 
possibly detects only X-ray luminous AGN, while less luminous and/or heavily obscured AGN are likely to remain undetected. 
\item The large difference in the detection rates found at the radio (34 per cent) and X-ray (7.9 per cent) wavelengths 
for our sample DOGs suggests the limitations of X-ray surveys. The deep radio observations are found more effective in unveiling 
the AGN population residing in obscured environments.    
\item We find that the median-stacked images at 400~MHz uGMRT and 1.5~GHz JVLA show a clear 
detection of radio emission in otherwise radio-undetected DOGs, suggesting them to possess faint 
radio emission that remained undetected in our observations. 
The average 1.5 GHz radio luminosity of the radio-undetected DOGs infers them to be powered by 
either low-luminosity AGN or high star-formation rate of 100~M$\odot$~yr$^{-1}$ or both.  
\item Our stacked images yield a clear detection of radio emission with SNR $\simeq$ 20 and 
noise-rms of 3.4~$\mu$Jy~beam$^{-1}$ at 400 MHz and 1.5~$\mu$Jy~beam$^{-1}$ at 1.5~GHz. 
 Therefore, our stacking analysis demonstrates that the planned deeper radio surveys from the SKA and its pathfinder would enable us to detect the radio source population 
that remained undetected in the current 400~MHz uGMRT and 1.5 GHz JVLA observations. 
\end{itemize}




%

\section*{Acknowledgements} 
 We thank the anonymous reviewer for useful suggestions that helped us to improve the manuscript.
AK, VS and SD acknowledge the support from the Physical Research Laboratory, Ahmedabad, funded by the Department of Space, Government of India. 
CHI and YW acknowledge the support of the Department of Atomic Energy, Government of India, under project no. 12-R\&D-TFR5.02-0700.
We thank the staff of GMRT who have made these observations possible. GMRT is run by the National Centre for Radio 
Astrophysics of the Tata Institute of Fundamental Research.
This work is based on observations made with the Spitzer Space Telescope, which is operated by the 
Jet Propulsion Laboratory, California Institute of Technology under NASA.
This paper makes use of software developed for Vera C. Rubin Observatory. We thank the Rubin Observatory for making their code available as free software at http://pipelines.lsst.io/.
This paper is based on data collected from the Subaru Telescope and retrieved from the HSC data archive system, which is operated by the Subaru Telescope and Astronomy Data Center (ADC) at NAOJ. Data analysis was in part carried out with the cooperation of Center for Computational Astrophysics (CfCA), NAOJ. 
\vspace{-1em}


%
%

\bibliography{RadioDOGPaper}{}

\begin{thebibliography}{}
\expandafter\ifx\csname natexlab\endcsname\relax\def\natexlab#1{#1}\fi

\bibitem[{{Aihara} {$et~al$.}(2022){Aihara}, {AlSayyad}, {Ando}, {Armstrong},
  {Bosch}, {Egami}, {Furusawa}, {Furusawa}, {Harasawa}, {Harikane}, {Hsieh},
  {Ikeda}, {Ito}, {Iwata}, {Kodama}, {Koike}, {Kokubo}, {Komiyama}, {Li},
  {Liang}, {Lin}, {Lupton}, {Lust}, {MacArthur}, {Mawatari}, {Mineo},
  {Miyatake}, {Miyazaki}, {More}, {Morishima}, {Murayama}, {Nakajima},
  {Nakata}, {Nishizawa}, {Oguri}, {Okabe}, {Okura}, {Ono}, {Osato}, {Ouchi},
  {Pan}, {Plazas Malag{\'o}n}, {Price}, {Reed}, {Rykoff}, {Shibuya},
  {Simunovic}, {Strauss}, {Sugimori}, {Suto}, {Suzuki}, {Takada}, {Takagi},
  {Takata}, {Takita}, {Tanaka}, {Tang}, {Taranu}, {Terai}, {Toba}, {Turner},
  {Uchiyama}, {Vijarnwannaluk}, {Waters}, {Yamada}, {Yamamoto}, \&
  {Yamashita}}]{Aihara22}
{Aihara}, H., {AlSayyad}, Y., {Ando}, M., {$et~al$.} 2022, \pasj,
  arXiv:2108.13045

\bibitem[{{An} \& {Baan}(2012)}]{An12}
{An}, T., \& {Baan}, W.~A. 2012, \apj, 760, 77

\bibitem[{{Becker} {$et~al$.}(1995){Becker}, {White}, \& {Helfand}}]{Becker95}
{Becker}, R.~H., {White}, R.~L., \& {Helfand}, D.~J. 1995, \apj, 450, 559

\bibitem[{{Chen} {$et~al$.}(2018){Chen}, {Brandt}, {Luo}, {Ranalli}, {Yang},
  {Alexander}, {Bauer}, {Kelson}, {Lacy}, {Nyland}, {Tozzi}, {Vito},
  {Cirasuolo}, {Gilli}, {Jarvis}, {Lehmer}, {Paolillo}, {Schneider}, {Shemmer},
  {Smail}, {Sun}, {Tanaka}, {Vaccari}, {Vignali}, {Xue}, {Banerji}, {Chow},
  {H{\"a}u{\ss}ler}, {Norris}, {Silverman}, \& {Trump}}]{Chen18}
{Chen}, C. T.~J., {Brandt}, W.~N., {Luo}, B., {$et~al$.} 2018, \mnras, 478,
  2132

\bibitem[{{Condon} {$et~al$.}(2002){Condon}, {Cotton}, \&
  {Broderick}}]{Condon02}
{Condon}, J.~J., {Cotton}, W.~D., \& {Broderick}, J.~J. 2002, \aj, 124, 675

\bibitem[{{Condon} {$et~al$.}(1998){Condon}, {Cotton}, {Greisen}, {Yin},
  {Perley}, {Taylor}, \& {Broderick}}]{Condon98}
{Condon}, J.~J., {Cotton}, W.~D., {Greisen}, E.~W., {$et~al$.} 1998, \aj, 115,
  1693

\bibitem[{{Cool} {$et~al$.}(2013){Cool}, {Moustakas}, {Blanton}, {Burles},
  {Coil}, {Eisenstein}, {Wong}, {Zhu}, {Aird}, {Bernstein}, {Bolton}, {Hogg},
  \& {Mendez}}]{Cool13}
{Cool}, R.~J., {Moustakas}, J., {Blanton}, M.~R., {$et~al$.} 2013, \apj, 767,
  118

\bibitem[{{Dey} {$et~al$.}(2008){Dey}, {Soifer}, {Desai}, {Brand}, {Le Floc'h},
  {Brown}, {Jannuzi}, {Armus}, {Bussmann}, {Brodwin}, {Bian}, {Eisenhardt},
  {Higdon}, {Weedman}, \& {Willner}}]{Dey08}
{Dey}, A., {Soifer}, B.~T., {Desai}, V., {$et~al$.} 2008, \apj, 677, 943

\bibitem[{{Donley} {$et~al$.}(2012){Donley}, {Koekemoer}, {Brusa}, {Capak},
  {Cardamone}, {Civano}, {Ilbert}, {Impey}, {Kartaltepe}, {Miyaji}, {Salvato},
  {Sanders}, {Trump}, \& {Zamorani}}]{Donley12}
{Donley}, J.~L., {Koekemoer}, A.~M., {Brusa}, M., {$et~al$.} 2012, \apj, 748,
  142

\bibitem[{{Fan} {$et~al$.}(2020){Fan}, {Chen}, {An}, {Xie}, {Han}, {Knudsen},
  \& {Yang}}]{Fan20}
{Fan}, L., {Chen}, W., {An}, T., {$et~al$.} 2020, \apjl, 905, L32

\bibitem[{{Farrah} {$et~al$.}(2008){Farrah}, {Lonsdale}, {Weedman}, {Spoon},
  {Rowan-Robinson}, {Polletta}, {Oliver}, {Houck}, \& {Smith}}]{Farrah08}
{Farrah}, D., {Lonsdale}, C.~J., {Weedman}, D.~W., {$et~al$.} 2008, \apj, 677,
  957

\bibitem[{{Farrah} {$et~al$.}(2017){Farrah}, {Petty}, {Connolly}, {Blain},
  {Efstathiou}, {Lacy}, {Stern}, {Lake}, {Jarrett}, {Bridge}, {Eisenhardt},
  {Benford}, {Jones}, {Tsai}, {Assef}, {Wu}, \& {Moustakas}}]{Farrah17}
{Farrah}, D., {Petty}, S., {Connolly}, B., {$et~al$.} 2017, \apj, 844, 106

\bibitem[{{Fiore} {$et~al$.}(2008){Fiore}, {Grazian}, {Santini}, {Puccetti},
  {Brusa}, {Feruglio}, {Fontana}, {Giallongo}, {Comastri}, {Gruppioni},
  {Pozzi}, {Zamorani}, \& {Vignali}}]{Fiore08}
{Fiore}, F., {Grazian}, A., {Santini}, P., {$et~al$.} 2008, \apj, 672, 94

\bibitem[{{Fiore} {$et~al$.}(2009){Fiore}, {Puccetti}, {Brusa}, {Salvato},
  {Zamorani}, {Aldcroft}, {Aussel}, {Brunner}, {Capak}, {Cappelluti}, {Civano},
  {Comastri}, {Elvis}, {Feruglio}, {Finoguenov}, {Fruscione}, {Gilli},
  {Hasinger}, {Koekemoer}, {Kartaltepe}, {Ilbert}, {Impey}, {Le Floc'h},
  {Lilly}, {Mainieri}, {Martinez-Sansigre}, {McCracken}, {Menci}, {Merloni},
  {Miyaji}, {Sanders}, {Sargent}, {Schinnerer}, {Scoville}, {Silverman},
  {Smolcic}, {Steffen}, {Santini}, {Taniguchi}, {Thompson}, {Trump}, {Vignali},
  {Urry}, \& {Yan}}]{Fiore09}
{Fiore}, F., {Puccetti}, S., {Brusa}, M., {$et~al$.} 2009, \apj, 693, 447

\bibitem[{{Frey} {$et~al$.}(2016){Frey}, {Paragi}, {Gab{\'a}nyi}, \&
  {An}}]{Frey16}
{Frey}, S., {Paragi}, Z., {Gab{\'a}nyi}, K.~{\'E}., \& {An}, T. 2016, \mnras,
  455, 2058

\bibitem[{{Gab{\'a}nyi} {$et~al$.}(2021){Gab{\'a}nyi}, {Frey}, \&
  {Perger}}]{Gabanyi21}
{Gab{\'a}nyi}, K.~{\'E}., {Frey}, S., \& {Perger}, K. 2021, \mnras, 506, 3641

\bibitem[{{Garilli} {$et~al$.}(2014){Garilli}, {Guzzo}, {Scodeggio},
  {Bolzonella}, {Abbas}, {Adami}, {Arnouts}, {Bel}, {Bottini}, {Branchini},
  {Cappi}, {Coupon}, {Cucciati}, {Davidzon}, {De Lucia}, {de la Torre},
  {Franzetti}, {Fritz}, {Fumana}, {Granett}, {Ilbert}, {Iovino}, {Krywult}, {Le
  Brun}, {Le F{\`e}vre}, {Maccagni}, {Ma{\l}ek}, {Marulli}, {McCracken},
  {Paioro}, {Polletta}, {Pollo}, {Schlagenhaufer}, {Tasca}, {Tojeiro},
  {Vergani}, {Zamorani}, {Zanichelli}, {Burden}, {Di Porto}, {Marchetti},
  {Marinoni}, {Mellier}, {Moscardini}, {Nichol}, {Peacock}, {Percival},
  {Phleps}, \& {Wolk}}]{Garilli14}
{Garilli}, B., {Guzzo}, L., {Scodeggio}, M., {$et~al$.} 2014, \aap, 562, A23

\bibitem[{{Hale} {$et~al$.}(2019){Hale}, {Williams}, {Jarvis}, {Hardcastle},
  {Morabito}, {Shimwell}, {Tasse}, {Best}, {Harwood}, {Heywood}, {Prandoni},
  {R{\"o}ttgering}, {Sabater}, {Smith}, \& {van Weeren}}]{Hale19}
{Hale}, C.~L., {Williams}, W., {Jarvis}, M.~J., {$et~al$.} 2019, \aap, 622, A4

\bibitem[{{Helfand} {$et~al$.}(2015){Helfand}, {White}, \&
  {Becker}}]{Helfand15}
{Helfand}, D.~J., {White}, R.~L., \& {Becker}, R.~H. 2015, \apj, 801, 26

\bibitem[{{Heywood} {$et~al$.}(2020){Heywood}, {Hale}, {Jarvis}, {Makhathini},
  {Peters}, {Sebokolodi}, \& {Smirnov}}]{Heywood20}
{Heywood}, I., {Hale}, C.~L., {Jarvis}, M.~J., {$et~al$.} 2020, \mnras, 496,
  3469

\bibitem[{{Heywood} {$et~al$.}(2022){Heywood}, {Jarvis}, {Hale}, {Whittam},
  {Bester}, {Hugo}, {Kenyon}, {Prescott}, {Smirnov}, {Tasse}, {Afonso}, {Best},
  {Collier}, {Deane}, {Frank}, {Hardcastle}, {Knowles}, {Maddox}, {Murphy},
  {Prandoni}, {Randriamampandry}, {Santos}, {Sekhar}, {Tabatabaei}, {Taylor},
  \& {Thorat}}]{Heywood22}
{Heywood}, I., {Jarvis}, M.~J., {Hale}, C.~L., {$et~al$.} 2022, \mnras, 509,
  2150

\bibitem[{{Hopkins} {$et~al$.}(2006){Hopkins}, {Hernquist}, {Cox}, {Di Matteo},
  {Robertson}, \& {Springel}}]{Hopkins06}
{Hopkins}, P.~F., {Hernquist}, L., {Cox}, T.~J., {$et~al$.} 2006, \apjs, 163, 1

\bibitem[{{Hopkins} {$et~al$.}(2008){Hopkins}, {Hernquist}, {Cox}, \&
  {Kere{\v{s}}}}]{Hopkins08}
{Hopkins}, P.~F., {Hernquist}, L., {Cox}, T.~J., \& {Kere{\v{s}}}, D. 2008,
  \apjs, 175, 356

\bibitem[{{Ishwara-Chandra} {$et~al$.}(2020){Ishwara-Chandra}, {Taylor},
  {Green}, {Stil}, {Vaccari}, \& {Ocran}}]{Ishwar20}
{Ishwara-Chandra}, C.~H., {Taylor}, A.~R., {Green}, D.~A., {$et~al$.} 2020,
  \mnras, 497, 5383

\bibitem[{{Kauffmann} \& {Haehnelt}(2000)}]{Kauffmann2000}
{Kauffmann}, G., \& {Haehnelt}, M. 2000, \mnras, 311, 576

\bibitem[{{Lacy} \& {Sajina}(2020)}]{Lacy20b}
{Lacy}, M., \& {Sajina}, A. 2020, Nature Astronomy, 4, 352

\bibitem[{{Lacy} {$et~al$.}(2004){Lacy}, {Storrie-Lombardi}, {Sajina},
  {Appleton}, {Armus}, {Chapman}, {Choi}, {Fadda}, {Fang}, {Frayer},
  {Heinrichsen}, {Helou}, {Im}, {Marleau}, {Masci}, {Shupe}, {Soifer},
  {Surace}, {Teplitz}, {Wilson}, \& {Yan}}]{Lacy04}
{Lacy}, M., {Storrie-Lombardi}, L.~J., {Sajina}, A., {$et~al$.} 2004, \apjs,
  154, 166

\bibitem[{{Lacy} {$et~al$.}(2020){Lacy}, {Baum}, {Chandler}, {Chatterjee},
  {Clarke}, {Deustua}, {English}, {Farnes}, {Gaensler}, {Gugliucci},
  {Hallinan}, {Kent}, {Kimball}, {Law}, {Lazio}, {Marvil}, {Mao}, {Medlin},
  {Mooley}, {Murphy}, {Myers}, {Osten}, {Richards}, {Rosolowsky}, {Rudnick},
  {Schinzel}, {Sivakoff}, {Sjouwerman}, {Taylor}, {White}, {Wrobel},
  {Andernach}, {Beasley}, {Berger}, {Bhatnager}, {Birkinshaw}, {Bower},
  {Brandt}, {Brown}, {Burke-Spolaor}, {Butler}, {Comerford}, {Demorest}, {Fu},
  {Giacintucci}, {Golap}, {G{\"u}th}, {Hales}, {Hiriart}, {Hodge}, {Horesh},
  {Ivezi{\'c}}, {Jarvis}, {Kamble}, {Kassim}, {Liu}, {Loinard}, {Lyons},
  {Masters}, {Mezcua}, {Moellenbrock}, {Mroczkowski}, {Nyland}, {O'Dea},
  {O'Sullivan}, {Peters}, {Radford}, {Rao}, {Robnett}, {Salcido}, {Shen},
  {Sobotka}, {Witz}, {Vaccari}, {van Weeren}, {Vargas}, {Williams}, \&
  {Yoon}}]{Lacy20}
{Lacy}, M., {Baum}, S.~A., {Chandler}, C.~J., {$et~al$.} 2020, \pasp, 132,
  035001

\bibitem[{{Lansbury} {$et~al$.}(2020){Lansbury}, {Banerji}, {Fabian}, \&
  {Temple}}]{Lansbury20}
{Lansbury}, G.~B., {Banerji}, M., {Fabian}, A.~C., \& {Temple}, M.~J. 2020,
  \mnras, 495, 2652

\bibitem[{{Le F{\`e}vre} {$et~al$.}(2013){Le F{\`e}vre}, {Cassata}, {Cucciati},
  {Garilli}, {Ilbert}, {Le Brun}, {Maccagni}, {Moreau}, {Scodeggio}, {Tresse},
  {Zamorani}, {Adami}, {Arnouts}, {Bardelli}, {Bolzonella}, {Bondi},
  {Bongiorno}, {Bottini}, {Cappi}, {Charlot}, {Ciliegi}, {Contini}, {de la
  Torre}, {Foucaud}, {Franzetti}, {Gavignaud}, {Guzzo}, {Iovino}, {Lemaux},
  {L{\'o}pez-Sanjuan}, {McCracken}, {Marano}, {Marinoni}, {Mazure}, {Mellier},
  {Merighi}, {Merluzzi}, {Paltani}, {Pell{\`o}}, {Pollo}, {Pozzetti},
  {Scaramella}, {Tasca}, {Vergani}, {Vettolani}, {Zanichelli}, \&
  {Zucca}}]{LeFevre13}
{Le F{\`e}vre}, O., {Cassata}, P., {Cucciati}, O., {$et~al$.} 2013, \aap, 559,
  A14

\bibitem[{{Lonsdale} {$et~al$.}(2003){Lonsdale}, {Smith}, {Rowan-Robinson},
  {Surace}, {Shupe}, {Xu}, {Oliver}, {Padgett}, {Fang}, {Conrow},
  {Franceschini}, {Gautier}, {Griffin}, {Hacking}, {Masci}, {Morrison},
  {O'Linger}, {Owen}, {P{\'e}rez-Fournon}, {Pierre}, {Puetter}, {Stacey},
  {Castro}, {Polletta}, {Farrah}, {Jarrett}, {Frayer}, {Siana}, {Babbedge},
  {Dye}, {Fox}, {Gonzalez-Solares}, {Salaman}, {Berta}, {Condon}, {Dole}, \&
  {Serjeant}}]{Lonsdale03}
{Lonsdale}, C.~J., {Smith}, H.~E., {Rowan-Robinson}, M., {$et~al$.} 2003,
  \pasp, 115, 897

\bibitem[{{Lonsdale} {$et~al$.}(2015){Lonsdale}, {Lacy}, {Kimball}, {Blain},
  {Whittle}, {Wilkes}, {Stern}, {Condon}, {Kim}, {Assef}, {Tsai}, {Efstathiou},
  {Jones}, {Eisenhardt}, {Bridge}, {Wu}, {Lonsdale}, {Jones}, {Jarrett}, \&
  {Smith}}]{Lonsdale15}
{Lonsdale}, C.~J., {Lacy}, M., {Kimball}, A.~E., {$et~al$.} 2015, \apj, 813, 45

\bibitem[{{Melbourne} {$et~al$.}(2012){Melbourne}, {Soifer}, {Desai}, {Pope},
  {Armus}, {Dey}, {Bussmann}, {Jannuzi}, \& {Alberts}}]{Melbourne12}
{Melbourne}, J., {Soifer}, B.~T., {Desai}, V., {$et~al$.} 2012, \aj, 143, 125

\bibitem[{{Mohan} \& {Rafferty}(2015)}]{Mohan15}
{Mohan}, N., \& {Rafferty}, D. 2015, {PyBDSF: Python Blob Detection and Source
  Finder}, ascl:1502.007

\bibitem[{{Noboriguchi} {$et~al$.}(2019){Noboriguchi}, {Nagao}, {Toba},
  {Niida}, {Kajisawa}, {Onoue}, {Matsuoka}, {Yamashita}, {Chang}, {Kawaguchi},
  {Komiyama}, {Nobuhara}, {Terashima}, \& {Ueda}}]{Noboriguchi19}
{Noboriguchi}, A., {Nagao}, T., {Toba}, Y., {$et~al$.} 2019, \apj, 876, 132

\bibitem[{{Norris} {$et~al$.}(2021){Norris}, {Marvil}, {Collier},
  {Kapi{\'n}ska}, {O'Brien}, {Rudnick}, {Andernach}, {Asorey}, {Brown},
  {Br{\"u}ggen}, {Crawford}, {English}, {Rahman}, {Filipovi{\'c}}, {Gordon},
  {G{\"u}rkan}, {Hale}, {Hopkins}, {Huynh}, {HyeongHan}, {James Jee},
  {Koribalski}, {Lenc}, {Luken}, {Parkinson}, {Prandoni}, {Raja}, {Reiprich},
  {Riseley}, {Shabala}, {Sheil}, {Vernstrom}, {Whiting}, {Allison}, {Anderson},
  {Ball}, {Bell}, {Bunton}, {Galvin}, {Gupta}, {Hotan}, {Jacka}, {Macgregor},
  {Mahony}, {Maio}, {Moss}, {Pandey-Pommier}, \& {Voronkov}}]{Norris21}
{Norris}, R.~P., {Marvil}, J., {Collier}, J.~D., {$et~al$.} 2021, \pasa, 38,
  e046

\bibitem[{{O'Dea} \& {Saikia}(2021)}]{ODea21}
{O'Dea}, C.~P., \& {Saikia}, D.~J. 2021, \aapr, 29, 3

\bibitem[{{Patil} {$et~al$.}(2020){Patil}, {Nyland}, {Whittle}, {Lonsdale},
  {Lacy}, {Lonsdale}, {Mukherjee}, {Trapp}, {Kimball}, {Lanz}, {Wilkes},
  {Blain}, {Harwood}, {Efstathiou}, \& {Vlahakis}}]{Patil20}
{Patil}, P., {Nyland}, K., {Whittle}, M., {$et~al$.} 2020, \apj, 896, 18

\bibitem[{{Ricci} {$et~al$.}(2015){Ricci}, {Ueda}, {Koss}, {Trakhtenbrot},
  {Bauer}, \& {Gandhi}}]{Ricci15}
{Ricci}, C., {Ueda}, Y., {Koss}, M.~J., {$et~al$.} 2015, \apjl, 815, L13

\bibitem[{{Ricci} {$et~al$.}(2021){Ricci}, {Privon}, {Pfeifle}, {Armus},
  {Iwasawa}, {Torres-Alb{\`a}}, {Satyapal}, {Bauer}, {Treister}, {Ho}, {Aalto},
  {Ar{\'e}valo}, {Barcos-Mu{\~n}oz}, {Charmandaris}, {Diaz-Santos}, {Evans},
  {Gao}, {Inami}, {Koss}, {Lansbury}, {Linden}, {Medling}, {Sanders}, {Song},
  {Stern}, {U}, {Ueda}, \& {Yamada}}]{Ricci21}
{Ricci}, C., {Privon}, G.~C., {Pfeifle}, R.~W., {$et~al$.} 2021, \mnras, 506,
  5935

\bibitem[{{Riguccini} {$et~al$.}(2019){Riguccini}, {Treister},
  {Men{\'e}ndez-Delmestre}, {Cardamone}, {Civano}, {Gon{\c{c}}alves},
  {Hasinger}, {Koekemoer}, {Lanzuisi}, {Le Floc'h}, {Lusso}, {Lutz},
  {Marchesi}, {Miyaji}, {Pozzi}, {Ricci}, {Rodighiero}, {Salvato}, {Sanders},
  {Schawinski}, \& {Suh}}]{Riguccini19}
{Riguccini}, L.~A., {Treister}, E., {Men{\'e}ndez-Delmestre}, K., {$et~al$.}
  2019, \aj, 157, 233

\bibitem[{{Rowan-Robinson}(2000)}]{Rowan-Robinson2000}
{Rowan-Robinson}, M. 2000, \mnras, 316, 885

\bibitem[{{Schuldt} {$et~al$.}(2021){Schuldt}, {Suyu}, {Ca{\~n}ameras},
  {Taubenberger}, {Meinhardt}, {Leal-Taix{\'e}}, \& {Hsieh}}]{Schuldt21}
{Schuldt}, S., {Suyu}, S.~H., {Ca{\~n}ameras}, R., {$et~al$.} 2021, \aap, 651,
  A55

\bibitem[{{Shanks} {$et~al$.}(2021){Shanks}, {Ansarinejad}, {Bielby},
  {Heywood}, {Metcalfe}, \& {Wang}}]{Shanks21}
{Shanks}, T., {Ansarinejad}, B., {Bielby}, R.~M., {$et~al$.} 2021, \mnras, 505,
  1509

\bibitem[{{Singh} {$et~al$.}(2018){Singh}, {Chand}, {Ishwara-Chandra}, \&
  {Kharb}}]{Singh18}
{Singh}, V., {Chand}, H., {Ishwara-Chandra}, C.~H., \& {Kharb}, P. 2018, in
  Revisiting Narrow-Line Seyfert 1 Galaxies and their Place in the Universe, 29

\bibitem[{{Singh} {$et~al$.}(2015){Singh}, {Ishwara-Chandra}, {Wadadekar},
  {Beelen}, \& {Kharb}}]{Singh15}
{Singh}, V., {Ishwara-Chandra}, C.~H., {Wadadekar}, Y., {Beelen}, A., \&
  {Kharb}, P. 2015, \mnras, 446, 599

\bibitem[{{Stern} {$et~al$.}(2005){Stern}, {Eisenhardt}, {Gorjian}, {Kochanek},
  {Caldwell}, {Eisenstein}, {Brodwin}, {Brown}, {Cool}, {Dey}, {Green},
  {Jannuzi}, {Murray}, {Pahre}, \& {Willner}}]{Stern05}
{Stern}, D., {Eisenhardt}, P., {Gorjian}, V., {$et~al$.} 2005, \apj, 631, 163

\bibitem[{{Tanaka} {$et~al$.}(2018){Tanaka}, {Coupon}, {Hsieh}, {Mineo},
  {Nishizawa}, {Speagle}, {Furusawa}, {Miyazaki}, \& {Murayama}}]{Tanaka18}
{Tanaka}, M., {Coupon}, J., {Hsieh}, B.-C., {$et~al$.} 2018, \pasj, 70, S9

\bibitem[{{Toba} {$et~al$.}(2015){Toba}, {Nagao}, {Strauss}, {Aoki}, {Goto},
  {Imanishi}, {Kawaguchi}, {Terashima}, {Ueda}, {Bosch}, {Bundy}, {Doi},
  {Inami}, {Komiyama}, {Lupton}, {Matsuhara}, {Matsuoka}, {Miyazaki},
  {Morokuma}, {Nakata}, {Oi}, {Onoue}, {Oyabu}, {Price}, {Tait}, {Takata},
  {Tanaka}, {Terai}, {Turner}, {Uchida}, {Usuda}, {Utsumi}, {Yamada}, \&
  {Wang}}]{Toba15}
{Toba}, Y., {Nagao}, T., {Strauss}, M.~A., {$et~al$.} 2015, \pasj, 67, 86

\bibitem[{{Truebenbach} \& {Darling}(2017)}]{Truebenbach17}
{Truebenbach}, A.~E., \& {Darling}, J. 2017, \mnras, 468, 196

\bibitem[{{Tsai} {$et~al$.}(2015){Tsai}, {Eisenhardt}, {Wu}, {Stern}, {Assef},
  {Blain}, {Bridge}, {Benford}, {Cutri}, {Griffith}, {Jarrett}, {Lonsdale},
  {Masci}, {Moustakas}, {Petty}, {Sayers}, {Stanford}, {Wright}, {Yan},
  {Leisawitz}, {Liu}, {Mainzer}, {McLean}, {Padgett}, {Skrutskie}, {Gelino},
  {Beichman}, \& {Juneau}}]{Tsai15}
{Tsai}, C.-W., {Eisenhardt}, P. R.~M., {Wu}, J., {$et~al$.} 2015, \apj, 805, 90

\bibitem[{{White} {$et~al$.}(2007){White}, {Helfand}, {Becker}, {Glikman}, \&
  {de Vries}}]{White07}
{White}, R.~L., {Helfand}, D.~J., {Becker}, R.~H., {Glikman}, E., \& {de
  Vries}, W. 2007, \apj, 654, 99

\bibitem[{{Willott} {$et~al$.}(2003){Willott}, {Rawlings}, {Jarvis}, \&
  {Blundell}}]{Willott03}
{Willott}, C.~J., {Rawlings}, S., {Jarvis}, M.~J., \& {Blundell}, K.~M. 2003,
  \mnras, 339, 173

\bibitem[{{Yutani} {$et~al$.}(2022){Yutani}, {Toba}, {Baba}, \&
  {Wada}}]{Yutani22}
{Yutani}, N., {Toba}, Y., {Baba}, S., \& {Wada}, K. 2022, arXiv e-prints,
  arXiv:2205.00567

\end{thebibliography}

\end{document}